\documentclass{elsarticle}
\usepackage{amsmath}
\usepackage{amssymb}
\usepackage{natbib}

\usepackage{graphicx}
\usepackage{csquotes}
\usepackage{gensymb}

\biboptions{sort&compress}

\begin{document}

\title{Re-examining the Role of Nuclear Fusion in a Renewables-Based Energy Mix}

\author[1]{T. E. G. Nicholas\corref{cor1}}
\ead{thomas.nicholas@york.ac.uk}
\cortext[cor1]{Corresponding author}

\address[1]{
	York Plasma Institute, 
	Department of Physics, University of York, 
	Heslington, York YO10 5DD, UK
}

\author[2]{T. P. Davis}
\address[2]{
	Department of Materials, University of Oxford, 
	Parks Road, Oxford, OX1 3PH
}

\author[1]{F. Federici}

\author[3]{J. E. Leland}
\address[3]{
	Department of Electrical Engineering and Electronics, University of Liverpool, 
	Liverpool, L69 3GJ, UK
}

\author[1]{B. S. Patel}

\author[4]{C. Vincent}
\address[4]{
	Centre for Advanced Instrumentation, 
	Department of Physics, Durham University, 
	Durham DH1 3LS, UK
}

\author[1]{S. H. Ward}

\newpageafter{abstract}

\date{\today}
\begin{abstract}
Fusion energy is often regarded as a long-term solution to the world's energy needs.
However, even after solving the critical research challenges, engineering and materials science will still impose significant constraints on the characteristics of a fusion power plant.
Meanwhile, the global energy grid must transition to low-carbon sources by 2050 to prevent the worst effects of climate change.
We review three factors affecting fusion’s future trajectory: (1) the significant drop in the price of renewable energy, (2) the intermittency of renewable sources and implications for future energy grids, and (3) the recent proposition of intermediate-level nuclear waste as a product of fusion.
Within the scenario assumed by our premises, we find that while there remains a clear motivation to develop fusion power plants, this motivation is likely weakened by the time they become available.
We also conclude that most current fusion reactor designs do not take these factors into account and, to increase market penetration, fusion research should consider relaxed nuclear waste design criteria, raw material availability constraints and load-following designs with pulsed operation.
\end{abstract}


\maketitle

\section{Introduction}

Nuclear fusion is often assumed to be the preferred source of baseload energy in a far-future energy mix; i.e. that once the technology is demonstrated, fusion's advantages make it a clear choice for low-carbon energy generation - assuming it is cost-competitive \cite{Bustreo2019}.
However, the relative advantages and disadvantages of fusion as a long-term energy source are complex. Rather than assuming cost-competitive fusion would be  a clear choice, we instead argue the cost will be broadly similar to fission, then review fusion’s distinguishing features in the context of a post-carbon energy grid.
This allows us to consider which broad scenario would be required in order for fusion to play a significant role in future energy supply. 
This analysis differs from previous work (such as \cite{Cabal2017} and \cite{Anyaeji2017}) by including recent results of fusion materials research and implications of climate scenarios involving urgent decarbonisation with low-cost renewables.

\begin{figure}[ht]
\centering
\includegraphics[width=0.6\columnwidth]{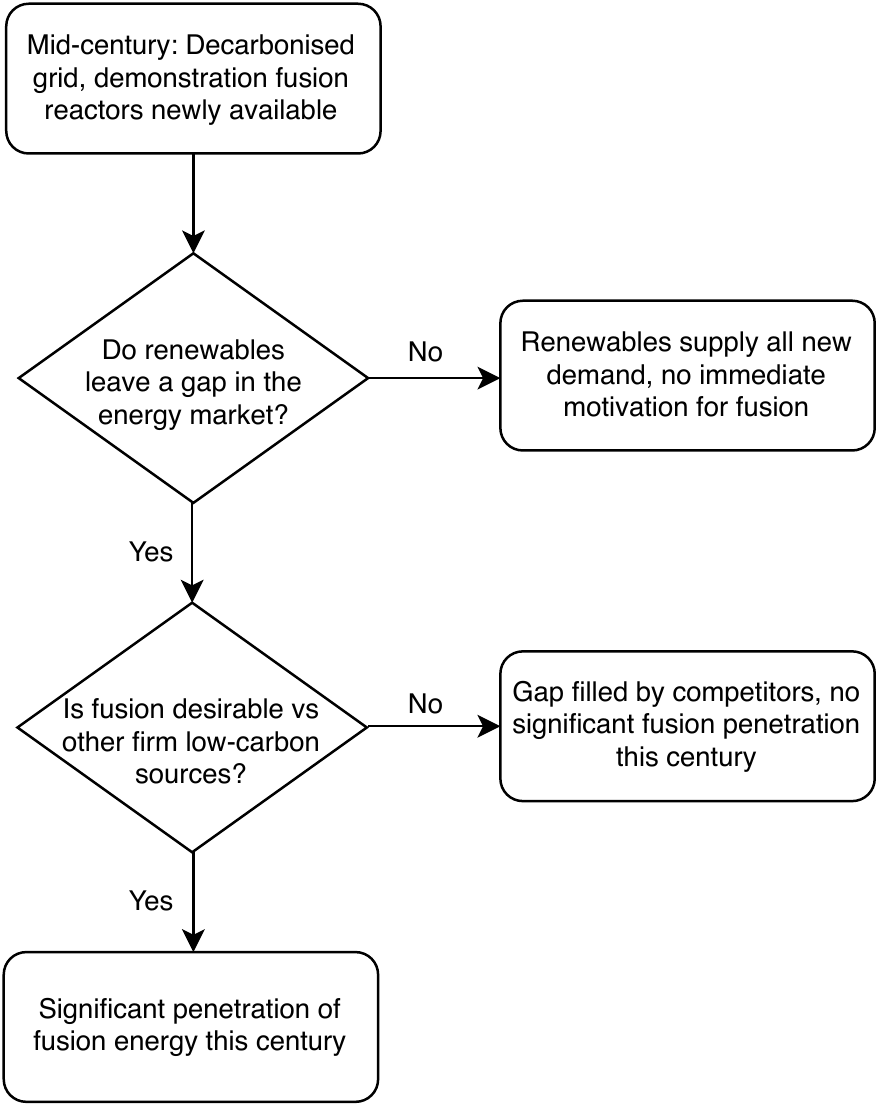}	
\caption{Scenario analysis for a post-carbon future. 	
In order for fusion to significantly penetrate the energy market this century, various conditions must be met within the market, policy, and fusion technology sphere.}	
\label{fig:argumentstructure}	
\end{figure}

After first giving the background context for fusion research, we justify some general premises describing relevant features of a post-carbon energy scenario and of future fusion power plants. We then examine advantages and disadvantages of fusion relative to other firm low-carbon sources, and summarise with recommendations for the fusion research program. The logical structure of our analysis is visualised in figure \ref{fig:argumentstructure}.

\section{Background}

Predating Keeling’s observation of rising atmospheric CO$_2$ levels\cite{Keeling1960}, research into commercial fusion power was originally motivated by its relative advantages over nuclear fission: non-proliferation, safety from meltdown, no long lived radioactive waste, a high power density, and abundant fuel.
This motivation was strengthened following the Three Mile Island and Chernobyl nuclear disasters, as degrading public sentiment toward nuclear power sources led to stagnation of new-build nuclear, opening a potential niche for commercial fusion \cite{Schneider2011}.

Whilst anthropogenic climate change has been acknowledged since the early 1970s, and the IPCC has issued reports since 1989, only very recently has focus started to shift towards the urgent need for complete decarbonisation by 2050 \cite{IPCC2018}.
Fusion’s value proposition has changed to meet this: the benefits of zero-carbon energy generation in particular are now emphasised, as well as arguments made for advantages over renewables, such as baseload supply, higher energy density, and geographical independence.

Currently the majority of global fusion efforts are working towards ITER, an intergovernmental-scale project to build the first tokamak aiming to demonstrate plasma energy breakeven.
ITER is set to operate at full power in 2035 and is projected to cost around \$22 billion\cite{Kramer2018}.
Most governmental fusion programs plan for an ITER-like demonstration fusion power plant (known as Demo) to be completed sometime after ITER.
Demo would demonstrate the necessary technologies, such as integrated tritium breeding, required to generate utility-scale baseload power.
Most governmental fusion programs (such as those proposed by China\cite{Zhuang2019} and Korea\cite{Kim2015}) use designs similar to the EU-DEMO1 design\cite{Federici2017}.
We acknowledge efforts to accelerate development by downsizing reactors using high-temperature superconducting (HTS) magnet technology\cite{Sorbom2015} but defer discussion of these until the end.

\section{Premises}

We first state six premises which could apply to the future energy mix or to fusion power technology in general.
The first two premises relate specifically to fusion technology: they arguably apply to all fusion devices, but are especially relevant to those operating with magnetically-confined fuel.
The latter four relate to the features of the future global energy mix. 

To be absolutely clear, we are assuming these premises to be true and exploring the resulting consequences; they are not intended as definite predictions, but are proposed as plausible and having implications worthy of consideration.

\subsection{Plasma physics challenges are solved}

We assume that plasma physics is understood well enough to design a commercial plant that can reliably operate in a scenario with sufficient plasma confinement for significant fusion power. We do not mean that knowledge of plasma confinement is enough to completely surpass the broad limits set by current confinement scaling laws\cite{Petty2008,ITERconf1999} (we exclude any possibility of ``tabletop fusion'') or significantly relaxed constraints on material properties of plasma-facing components\cite{Linsmeier2017}. This is currently not the case, and many major plasma physics challenges remain (as reviewed in the ITER physics basis \cite{Mukhovatov2007a}).  

This premise also excludes any power plants using a pure deuterium plasma or so-called aneutronic fuel mixtures, such as D-$^3$He, p-$^6$Li, or p-$^{11}$B.
This means a fusion power plant must use a deuterium-tritium fuel mixture and deal with the resulting high-energy neutrons.
While a fusion power plant without significant neutron activation would have enormous advantages, the physical arguments that this is likely impossible\cite{Rider1995, Rider1997} are well-known.

\subsection{Materials science challenges are solved}

We assume that the materials challenges\cite{Rowcliffe2018} are solved to the minimum extent to allow for operation of a fusion reactor using materials similar to those being considered for use in Demo\cite{Federici2017a}.
This includes plasma-facing materials that can withstand high heat loads\cite{Ueda2017} and structural materials\cite{Stork2014} that keep embrittlement and swelling issues at acceptable levels.
We do not assume that the materials used will not become nuclear waste through neutron activation.

\subsection{Energy grids will decarbonise without fusion}

Reduction of carbon dioxide emissions to net-zero by 2050\cite{IPCC2018} will be one of the most important factors steering the evolution of the energy sector.
With the EU committing in 2017 to targets which effectively require net-zero electricity supply by 2050 \cite{EU2017}, we assume a 95-100\% decarbonised power sector \cite{Jenkins2018b}. 
Hence we assume that the 2050 energy mix will comprise only zero-carbon technologies: fossil fuel power plants with carbon capture and storage (CCS), nuclear power plants, and renewables.

The EU fusion roadmap states ``[Demo] will be operational around 20 years after high power burning plasmas are demonstrated in ITER''\cite{Donne2018}, setting the earliest date for a commercial fusion plant to 2055.

Even after developing a demonstration reactor, the adoption of fusion plants must experience the so-called “valley of death” in which early plants have high capital costs and long build times, whilst not yet providing the optimised return on investment which would finance innovation \cite{Cardozo2016}.
This will be exacerbated by the challenges of developing new supply chains \cite{Surrey2019}.

Although this EU timeline is representative of governmental fusion research worldwide, there are also now several private companies aiming to develop commercial fusion power plants significantly earlier \cite{Sykes2018, Sorbom2015}.
However, due to the remaining physics, engineering and materials science challenges, this paper will assume that global energy supply is almost entirely decarbonised without contributions from fusion - whether publicly or privately funded.

This premise does not rely on decarbonisation being achieved by any specific date, only that decarbonisation occurs before widespread availability of commercial fusion power. The IPCC 1.5\degree C and EuroFusion timelines do not need to be interpreted as immovable dates in the context of this paper, only as a strong indication of events occurring in that order.

\subsection{Energy supply infrastructure is chosen primarily on the basis of monetary cost}

We assume that the choice of low-carbon energy sources is driven primarily by monetary cost and not military, geopolitical or other environmental reasons.

Given that a zero-carbon society is by definition more environmentally-conscious, we will also consider the possibility that alternative metrics, such as the Energy-Return-On-Invested (EROI), will be prioritised in future\cite{Carbajales-Dale2012}.

\subsection{Renewables will dominate decarbonised grids}

The cost of renewable electricity will continue to fall \cite{BEIS2016}. According to McKinsey: ``cheap renewable energy and batteries fundamentally reshape the electricity system […] by 2030 new-build renewables will outcompete existing fossil generation on energy cost in most countries'' \cite{McKinsey2019}.
Renewables are projected to reach high fractions (74\% globally by 2050) even without assuming significant carbon regulations \cite{McKinsey2019} or subsidies \cite{Lazard2019}.

Scenarios intended to meet stronger decarbonization targets display similarly high grid fractions: the IPCC\cite{IPCCSPM2018} state that ``In 1.5\degree C pathways with no or limited overshoot, renewables are projected to supply 70–85\% (interquartile range) of electricity in 2050 (high confidence).''
BP’s most recent projections envisage a scenario meeting net-zero by 2050 in which renewables collectively supply around 60\% of primary energy demand globally \cite{BP2020}.

The UK Committee on Climate Change similarly predicts that for meeting the 1.5$\degree$C target a renewable fraction of 78\% is desirable
A share of only 60\% is dubbed a ``cautious'' approach, because they predict that if higher fractions of renewables are possible, for example through long-distance interconnects, cheaper storage, or more demand-side management, an even larger share would likely reduce overall costs \cite{UKCCC2019}.

Therefore both the business-as-usual and high-decarbonisation scenarios still feature high fractions of renewables. 
Our assumption here is that any future decarbonised grid involves a high fraction of renewable generation.

Once decarbonisation has been achieved the energy mix will likely still evolve as innovation occurs. However, though the value of fusion should be continually reassessed as the market develops, if our premises remain true then our conclusions would still apply.

\subsection{Fusion energy will not be as cheap as renewable electricity}

Intuitively, fusion plants are economically similar to fission plants: large capital costs, resulting from generator turbines, cooling, concrete shielding and containment, high safety standards, nuclear licensing, decommissioning, and nuclear waste management; a high fraction of ongoing costs dedicated to operation and maintenance, components replacement and interest repayments; and relatively low fuel costs.
Magnetically-confined fusion also specifically requires large and expensive magnets.

Entler\cite{Entler2018} modelled the cost of a 1GW EU-Demo fusion power plant concept, finding a levelized cost of energy (LCOE) of 175\$/MWh with a direct capital cost (with contingency) of \$7.4 billion (all costs quoted in 2018 USD).
Fusion's LCOE is sensitive to the initial capital cost\cite{Sheffield2016}: a capital cost of \$3.9 billion for 1GW suggests a LCOE of 83\$/MWh, whereas using a similar capital cost to Entler’s Demo design (\$6.2 billion) yields a LCOE of 121\$/MWh.
Generally increasing the cost of the fusion technology by \$1 billion increases the cost of electricity by 16.5 \$/MWh. 
Setting the capital cost of the fusion-specific technology  to zero results in a LCOE of 72 \$/MWh, which borders current prices for renewables, but is clearly physically implausible.

Entler\cite{Entler2018} finds the effective operating and decommissioning costs alone for their fusion power plant concept come to 27 \$/MWh.
However, this assumes no long-lived radioactive waste, an assumption which we examine later in this paper.

Given that ITER and Hinkley Point C are both projected to cost over \$20 billion each\cite{Kramer2018, Haas2019}, we assume a cost over 100\$/MWh is more realistic.
For comparison, Lazard suggests that currently the LCOE of large solar PV is between 40-46\$/MWh, onshore wind between 29-56\$/MWh and offshore wind at 92\$/MWh \cite{Lazard2019}.

Therefore we assume both utility-scale solar and onshore wind to be significantly cheaper than fusion by the time fusion becomes commercially viable, and possibly indefinitely.

\section{Discussion}

\subsection{Can renewables handle it all?}

Several studies conclude that it is possible to meet national energy demands using only renewable sources (including geothermal and hydroelectric) \cite{Jacobson2018, Connolly2014, Ram2019, Ueckerdt2015, Child2019, Zappa2019}.
In specific cases it is actually cheaper than the cost of a business-as-usual scenario, with the caveat that energy grid composition is extremely region-specific and often includes long-distance energy transfer with international grid integration\cite{Pleissmann2017}.
For example Jacobson\cite{Jacobson2018} comprehensively analyses multiple scenarios where an all-renewables grid with storage provides load-following power at economically viable electricity prices.

A scenario in which the decarbonisation challenge is met entirely by renewables, without the need for ``baseload'' energy sources, would have no clear motivation to include fusion.
Assuming renewables and storage remain the cheapest option (including cost of energy storage) then, according to premise D, they would meet any subsequent increases in demand, without nuclear waste, safety or proliferation issues.

\subsection{Limits to penetration of renewables}

Renewables do have significant disadvantages that must be overcome for an all-renewable grid to be feasible, the most challenging of which is temporal intermittency of supply.
Some combination of energy storage, some other highly dispatchable energy source, and long-distance interconnects must be used to cope with fluctuating grid loads\cite{Gils2017}.
While \cite{Ziegler2019} shows that electrical storage may not be competitive if relied upon completely, utilising batteries (like scenario C in \cite{Jacobson2018}) can reduce costs drastically.
However, diversified energy mixes which include firm non-renewable backup are almost always cheaper overall\cite{Mileva2016,Sepulveda2018,Brick2016}.
Being energy-sparse, large-scale renewables deployment also entails significant land-use impacts\cite{Herendeen2019} and so is disadvantageous when land is high-value or energy demand is dense (e.g. for megacities).

\subsection{The desirability of fusion in this new context}

Instead, imagining a decarbonised scenario with high - but not total - renewable penetration by the second half of the 21st century, other low-carbon sources will need to fill the supply gap. 

When large-scale fusion deployment becomes plausible, the grid composition will be significantly different to today: new sources will have to work in tandem with the high renewables fraction.
Such a grid will value flexibility - as predicted for integrating both fission \cite{Cany2016} and gas with CCS \cite{Mechleri2017} with intermittent renewables.
If fusion cannot provide this service it may be excluded to alternative markets which require baseload.

The options for firm low-carbon generation are principally nuclear fission, nuclear fusion, and gas with carbon capture and storage.
As fusion will be competing with these sources, we now attempt to assess its relative merits, focusing on constraints likely to be imposed by the materials science and engineering challenges of developing a near-term fusion reactor design.

\subsubsection{Load-following}

Within a grid with large fractions of intermittent renewables, dispatchable energy sources that can match demand will lower overall system costs \cite{Mileva2016,Sepulveda2018,Brick2016,Jenkins2018a}.
This is expected for CCS\cite{Brouwer2015}, fission\cite{Jenkins2018b} and fusion\cite{Bustreo2019}.
For baseload plants, access may be improved by long-distance interconnects which smooth out supply variations\cite{Hamacher2013}, but even in that scenario dispatchability will still be highly valued.

This raises the question of technical challenges for a fusion power plant to load-follow rather than supplying baseload.
This has been answered briefly for the EU-Demo concept, concluding that it would be possible to reduce the fusion power by about 50\%\cite{Maisonnier2005}.
However, this is complicated by knock-on effects associated with reduced plasma power output\cite{Ward2015a}.

In any magnetic-confinement design there will be tension between load-following directly with the plasma power output and control of the plasma.
Whether devices operate in pulsed or steady state modes\cite{Federici2017,Donne2018}, real-time control systems must ``pilot'' them through a multidimensional parameter space, avoiding regions dangerous to the plasma confinement.
Load-following significantly increases the number of trajectories which must be understood and safely managed; for this reason the EU-Demo design is currently assuming a single operation scenario.
In practice this means running the same plasma pulse scenario repeatedly, with the same duration and energy output.

Alternatively, coupling a Demo-sized reactor in pulsed operation with a thermal battery between blanket and turbines would allow for smoothing of variable grid demand, but it would also significantly increase the cost \cite{Homonnay, Vanter2012}.
There are suggestions for instead co-generating hydrogen\cite{Sheffield2016}, though this would decrease overall energy efficiency.
It should be noted that this approach could also be taken with renewables or fission to allow for long term energy storage.

Similarly, running a fusion plant at full power to generate heat (see section 4.3.6) whilst simultaneously co-generating load-following electricity could improve the economics of load-following, as has been studied in the context of small modular fission reactors \cite{Locatelli2015}.

Fission has already demonstrated load following by directly altering the power output from the fission core\cite{Cany2016}, so this is not an area where fusion has any clear advantage.
While fission cores respond more slowly than tokamak plasmas due to decay heat and fuel poisoning constraints\cite{Ponciroli2017}, they could also couple to  thermal or hydrogen storage technology.

Of the low-carbon technologies, gas with CCS is the most suitable for load following; gas turbines are regularly used as ``peaker'' plants, and adding CCS doesn’t directly impinge on this capability \cite{Domenichini2013, Mechleri2017}.

It should also be noted that the financial model of a large fusion plant (high up-front costs, low fuel costs) is not in favour of lowering energy output.
If low demand can force a plant to reduce output, the relative opportunity cost of dormancy will become significant.
Sources where fuel prices are a large fraction of total costs benefit more when flexible operation is required\cite{Mechleri2017}.
It would therefore be unlikely for load-following fusion plants to be competitive economically unless there was some opposition to gas with CCS that disincentivised their construction (e.g. lack of subsidies, lack of available CO$_2$ storage, public opposition, infrastructure leakage preventing emissions compliance \cite{Alvarez2018}, etc.), or a use-scenario for fusion at scale that significantly improved its load-following economics (e.g. desalination or other co-generation \cite{Locatelli2015}).
Quantitative market modelling studies should examine this question more rigorously.

\subsubsection{Waste production}

One of the commonly stated advantages of fusion over fission is the misconception that it will not produce ``long-lived'' radioactive waste \cite{Gorley2015, Federici2017}.

In the 1980s the fusion materials community discussed methods to reduce the volume of long-lived radioactive waste generated by neutron activation\cite{Brager1985}.
In 1982 the U.S. Department of Energy decided to aim for significantly reduced volumes of high-level nuclear waste (the UK definitions of different waste classes are given in figure \ref{table:wasteclassification}) by limiting the radioactive lifetimes of fusion waste materials compared to fission materials \cite{Conn1983}.
This was implemented by introducing a ``low-activation’' or ``reduced-activation'' material criterion, which was defined as \cite{Gorley2015}:

\begin{figure}[ht]
  \centering
  \includegraphics[width=0.9\textwidth]{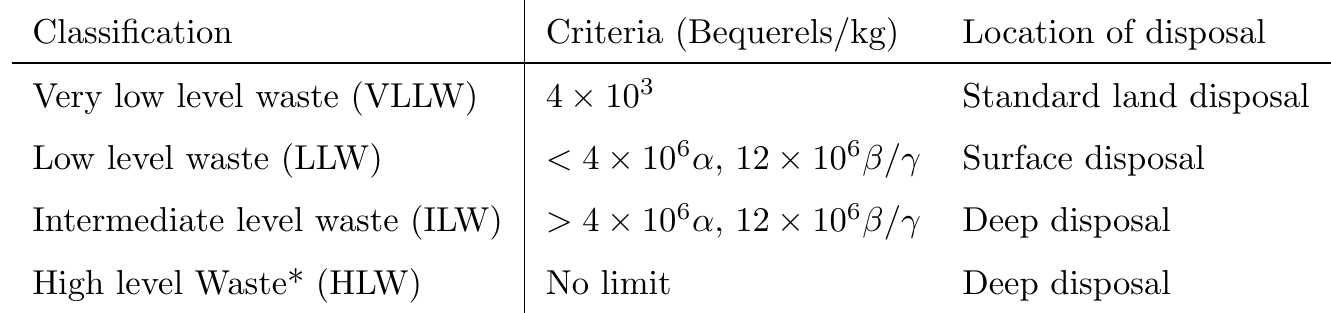}
  \caption{UK radioactive waste classification defined by the Nuclear Decommissioning Authority (\cite{NDA2017}: Radioactive Wastes in the UK: UK Radioactive Waste Inventory Report, Tech. Rep. (Nuclear Decommissioning Authority, 2017)). *High level nuclear waste is defined as waste in which the temperature may rise significantly as a result of their radioactivity, so this factor must be taken into account in the design of storage or disposal facilities.}
  \label{table:wasteclassification}
\end{figure}

\begin{displayquote}
“The materials selection for fusion energy’s nuclear waste production, after an initial $\sim$100 years removal from the reactor, can be disposed of in low-level waste repositories.”
\end{displayquote}

For this purpose reduced-activation (or low-activation) structural steels were designed, and neutronics modelling concluded these steels would meet the low-activation criterion and hence only be classified as low-level waste (LLW) \cite{Conn1984, Kohyama1996, Klueh2000, Gorley2015, Gilbert2018, Lindau2005}.
EUROFER97 reduced-activation steel is the leading fusion structural material that was designed by the Eurofusion effort \cite{Gaganidze2018} and has been chosen as the neutron-facing structural material in the EU-DEMO1 reactor design \cite{Federici2017}.

However, the latest research using the EU-DEMO1 design, by Gilbert et al. (2019) \cite{Gilbert2019} and Bailey et al. (2020) \cite{Bailey2020}, suggests that the EUROFER97 steel (used for the first wall, breeder blanket and divertor components) will always exceed the reduced-activation criterion, and hence be classed as intermediate-level waste (ILW). 

In fact, more intermediate-level waste (ILW) may be produced by fusion than by fission: the European Sodium-Cooled fission Fast Reactor (SCFR) design has a lower percentage of ILW per total reactor steel mass compared to the EU-DEMO1 design (for a similar power output)\cite{Reid2020}.
     
Fusion-specific structural steels will therefore be classified as intermediate-level waste (ILW) under the UK protocol, with half-lives of thousands of years for EUROFER97.
This is a significant volume of radioactive waste: recent EU-Demo designs require 1300-1500 metric tons of steel that will be strongly irradiated and thus become ILW\cite{Someya2015}.

Furthermore, Gilbert et al. \cite{Gilbert2019} indicate that any beryllium used in a Helium-Cooled Pebble Bed (HCPB) breeder blanket could exceed the reduced-activation criterion due to natural uranium impurities (which activate to become $^{239}$Pu and $^{241}$Am in trace amounts, which have extremely long half-lives).

Under current UK nuclear law, ILW requires geological disposal.
Therefore, the latest research indicates that fusion plants could produce nuclear waste which requires long-term subsurface disposal, similar to fission plants.
Although fusion won’t produce high-level waste (which requires active cooling), and also won’t produce radioisotopes with half-lives of $>100,000$ years, the need for deep geological disposal still weakens one of the main arguments for fusion over fission.

It is hard to avoid this: the elements responsible are either required for mechanical properties or are present as natural ore impurities in the tens of parts-per-million concentrations, the reduction of which  might not be technically or economically feasible.
It has been suggested that the structural materials could be recycled \cite{Pace2012}, however the social, technical and economic aspects of recycling hundreds of tons of nuclear waste for use in a new reactor have not been proven even in concept.

It should be noted that this is specific to UK regulatory law and would not necessarily apply internationally.
As explained in Gilbert’s paper\cite{Gilbert2019}, the levels of $^{14}$C that make EUROFER97 problematic in UK law are not a problem under French regulations, and conversely the levels of $^{94}$Nb which would be problematic for French regulators would not be an issue under UK regulatory law.
Additionally, the levels of both $^{14}$C and $^{94}$Nb do not constitute the need for geologic disposal in Japan at all.
It is therefore conceivable that the fusion community could lobby for a separate, internationally consistent categorisation for fusion structural waste based on arguing that inert steel poses a lower risk of biosphere penetration than waste from fission.
It is unclear how easy this would be and, if not possible, an alternative solution could be to relax the LLW criterion altogether and accept that fusion will generate ILW.

\subsubsection{Proliferation}

Despite valid concerns, fusion does still have an advantage over fission in terms of nuclear non-proliferation as safeguarding measures are potentially much easier.

Tritium is used in thermonuclear weapons but is of little use by itself since it requires fissionable material in the primary stage.
Goldston et al. \cite{Goldston2012} explored the possibility of using the $14.1$MeV fusion neutrons to generate fissionable material, but concluded this would be difficult to perform clandestinely and relatively straightforward to halt unilaterally, assuming appropriate detection safeguards were in place.
They also considered the potentially problematic possibility of boosting conventional atomic weapons with tritium. 

If we compare fusion to advanced fission concepts, the overall picture changes.
Generation IV fission reactors should, depending on the design, boast impressive non-proliferation credentials: advanced reprocessing and fast burnup of plutonium within the core of the reactor\cite{GenIVRoadmap2014} results in a smaller inventory of fissile material on site. 

The use of a thorium cycle is also argued to have better non-proliferation credentials.
However, the $^{233}$U that thorium produces through neutron capture is weaponisable, with the IAEA categorising $^{233}$U on the same basis as plutonium \cite{IAEA1980}.
Proponents conversely argue that $^{233}$U has a high $^{232}$U content - a strong gamma emitter - which therefore potentially makes it difficult to handle safely and easy to detect and safeguard against - though the extent to which this is true is disputed\cite{Hesketh2010}.

Whilst fusion is superior to traditional fission in terms of proliferation, it is harder to conclude its superiority over possible advanced fission concepts, which may be competitors by the time fusion enters the market.

\subsubsection{Energy Return On Invested}

Relative monetary cost of a technology does not completely reflect the associated environmental damages, and one suggestion for a more comprehensive metric has been the Energy Return On Invested (EROI) \cite{Carbajales-Dale2012}.

Estimates for EROI of energy sources vary in the literature, but broadly indicative comparisons of different current UK electricity sources have been made \cite{Raugei2016}.
The highest EROI found was hydroelectricity (170) followed by fission (87), wind (50-52), gas (41 - albeit without CCS) and PV (10-25).
Only one study has estimated the EROI of a fusion power plant (at $\sim~27$), but for a reactor design from 1975 \cite{White2000}.
We will therefore instead now try to place general bounds on the most optimistic possible value for fusion and other advanced nuclear technologies.

As fusion’s fuel is energetically dense and relatively abundant, then for either optimistic\cite{Lozada-Hidalgo2017} or routine\cite{Rae1978} estimates of enrichment costs, its fuel cycle still has a high EROI.
However, the EROI of the complete system is limited by the energy invested in the device required to unlock that potential energy
\begin{align}
\begin{split}
\text{EROI} &= \frac{\text{Energy Generated}}{\text{Energy Invested}} \\
            &= \frac{Pt_L}{Mt_L + FPt_L + I} \\
            &= \frac{1}{\frac{M}{P} + \frac{1}{\text{EROI}_F} + \frac{I}{Pt_L}},
\end{split}
\end{align}
where $P$ is reactor output power, $t_L$ is the plant lifetime, $M$ is the maintenance energy cost per unit time, $F$ is the energy cost of mining and producing fuel per unit of usable output energy, $I$ is the one-off energy cost of building and decommissioning the plant infrastructure, and $\text{EROI}_F=1/F$ is the EROI of the fuel alone.
Clearly, even if EROI$_F$ were infinite, system EROI does not tend to infinity, instead
\begin{equation}
    \text{EROI} \xrightarrow{} \frac{1}{\frac{M}{P} + \frac{I}{Pt_L}}.
\end{equation}

As discussed under premise F, any fusion power plant will require significant physical infrastructure, balance-of-plant, maintenance, and decommissioning.
Therefore, to roughly estimate fusion’s EROI, we assume the energy costs associated with plant construction, maintenance, and decommissioning (i.e. everything but the fuel procurement) are comparable to that of a fission plant of similar power output.
For fission pressurised water reactors (PWRs), the fuel cycle requires about half the total energy input of the whole technology life cycle \cite{Weissbach2013}, so even if the fuel were available for zero energy cost, the overall EROI would only approximately double to $\sim170$.
This represents an upper bound since we also know that the fusion reactor “island” has a minimum size \cite{Freidberg2015}, and that the power-generating plasma of a fusion plant has a lower volumetric power density than the core of a  fission plant ($\sim1.2$MW/m${}^3$ for EU-DEMO1 vs $\sim300$MW/m${}^3$ for a SCFR).

Although optimistic, this EROI is considerably higher than any existing (widely-scalable) technologies, implying a possible place for fusion in ecologically conscious grids.
Similar arguments apply to advanced fission fuels, where fissioning the $^{238}$U increases the EROI of the fission fuel by more than an order of magnitude, in a similar-sized reactor.
But the threefold advantage of fusion over wind (even including storage \cite{Carbajales-dale2014}) might be jeopardized by fusion-specific demand for elements which are energy-intensive to extract (e.g. beryllium, tungsten, rare-earth metals such as yttrium for superconducting magnets).

\subsubsection{Resource supply}

Whilst it is true that fusion has access to an abundant source of deuterium and lithium fuel,  it remains uncertain whether other reactor-relevant resources could become severe limiting factors.
Using Fasel’s \cite{Fasel2005} approach but with updated availability estimates \cite{JaksulaLi2020}, current accessible resources of terrestrial lithium could provide ~2800 years of fusion power.
But the increase in competition with other industries, notably batteries for energy storage and electric vehicles (EVs), could consume these reserves much faster - potentially within decades \cite{Bradshaw2011}.
As this usage of lithium is purely chemical - hence isotope-agnostic - an economy could be established whereby enriched $^6$Li is used solely by fusion reactors and ‘depleted’ $^7$Li used by the energy storage industry.
But this may not be likely, given that the EV industry is forecast to undergo extensive growth before fusion is commercialised \cite{IEA2019}.
Bradshaw\cite{Bradshaw2011} simply assumed a ``worst-case'' scenario, but the implications of lithium demand economics for fusion deserve more detailed study in future work.

Access to lithium in seawater would increase potential reserves for fusion and energy storage by several orders of magnitude, but the economic and environmental costs of processing the necessary quantities of seawater must be considered.
Considering purely the energy efficiency of extracting from the ocean\cite{Bardi2010}, the currently-projected EROI for lithium burned through fusion is only slightly $>1$.

Fission also faces limitations in the availability of economically extractable uranium.
The estimated uranium lifetime for the current light-water reactor (once-through) is $\sim100$ years at 2002 world nuclear electricity generation with known conventional uranium resources but increases beyond $\sim2000$ years if Generation IV nuclear reactors are considered\cite{RedBook2003}.
However, it should be made clear that these numbers are conservative \cite{Schneider2008}.
Additionally, thorium supply could theoretically be 3 times greater due to the greater abundance \cite{Schaffer2013}.

Unlike for fusion fuels, the EROI of extracting these fission fuels from seawater was evaluated as being $< 1$ when using contemporary technologies \cite{Bardi2010}, but research is ongoing\cite{Alexandratos2016}.
On the other hand, next-generation nuclear could alternatively use spent fuel and unenriched ${}^{238}$U which, along with the closed nature of their potential fuel cycles, means that it could greatly exceed 90 years\cite{Gabriel2013}.

Comparing this to gas with CCS, there exists an estimated upper limit on the amount of carbon storage available but this still allows for a high uptake of CCS until 2100 \cite{Budinis2018}.
These plants maintain a positive EROI despite the high energy cost of CCS, which results in a 6-20\% reduction in the energy output (depending on plant design). 

The beryllium used as a neutron multiplier in several fusion blanket  designs  is  also  a  serious concern.
As of 2018, the estimated identified world beryllium supply was $100,000$ metric tons, mined at a rate of 230 metric tons per year\cite{JaksulaBe2020}.
If a beryllium-based breeder blanket is selected for a DEMO reactor, a maximum of 200-300 reactors could be constructed\cite{Bradshaw2011}.
The beryllium cannot be fully recycled because it is transmuted during neutron multiplication, making it necessary to seek additional reserves or use an alternative.
Unfortunately the choices for neutron multipliers are fundamentally limited by nuclear physics, which leaves one of the only other options as lead, which must be used in liquid eutectic form and thus creates additional operational problems \cite{Malang2011,Abdou2015,Jun2020}. 

Magnetically-confined fusion schemes face another potential resource shortage in the liquid helium coolant required for the superconducting magnets\cite{Bruzzone2018}.
The total helium inventory contained in the number of fusion DEMO reactors needed to supply 30\% of global energy demand is around 2\% of the global helium resource, and helium has an exponential reserve index of around 100 years.
Bradshaw \cite{Bradshaw2013} concludes that even accounting for the helium directly produced by the D-T fusion reaction, fusion as a non-sustainable consumer of helium would exacerbate an already critical supply situation.
Substitution is challenging because of the low molecular mass required.
This constraint could be managed with foresight, active management of the resource supply chain and geopolitical cooperation, but large uncertainties remain.
However, Bradshaw also calculates that even energetically-costly extraction of the required helium from air would only lower Fusion’s EROI by around 1\%.

Renewables and gas with CCS also face material shortages of crucial elements, notably:  indium, gallium, and silver for photovoltaics\cite{Tao2011}, neodymium for wind, and nickel and molybdenum for CCS\cite{Kleijn2011, Bradshaw2011}.
It should be noted though that all these cases are limited by electrical or chemical processes, rather than nuclear reactions, and so there is arguably more potential for research into alternatives or recycling.

It must also be noted that all figures representing the amount of extractable material are inherently conservative and subject to revision. This can be seen in the increase in lithium resources from 25.5 to 80Mt between 2010 and 2020 \cite{JaksulaLi2020}, as the increase in lithium demand intensified both the search for new reserves and the development of new extraction methods intensified also.

It is therefore clear that fusion has an advantage over current generation fission in terms of fuel availability in the near term, but this advantage may not extend over advanced fission concepts.

\subsubsection{Can fusion supply heat?}

Heat at different temperatures is required for industrial and domestic use, and it comprises a significant fraction of global energy usage\cite{IEO2019, Thompson2013}.
An analysis of the 14 top greenhouse-gas-emitting industries in the United States (37\% of the total industry energy demand\cite{McMillan2018}) outlined the supply composition of heat of varying temperatures: 0.6\%, for low temperature ($<100\degree$C); 60.1\%, for medium temperature ($100-400\degree$C); and 39.3\% for high temperature ($>400\degree$C)\cite{McMillan2019a}.
Uses in the upper tier are then further stratified based on the required temperature.

Fusion can provide low and medium-temperature heat via the primary coolant (a maximum temperature of $\sim300\degree$C in water cooled designs\cite{Tobita2018}) or after the first high-pressure expansion stage (up to $700$C with liquid metal and helium coolants\cite{Abdou2015}), reaching 86\% of the total market demand.
Temperatures above this ($>800\degree$C) are technologically unattainable\cite{Abdou2015} for both fission and fusion designs.
As it stands, gas with CCS would fill this gap.

Since most heat demand can be supplied electrically, renewables could compete with fusion to provide low- and medium-temperature heat\cite{Wilson2013}.
However it must be noted that some renewables (e.g. solar) provide significantly less power during winter months when domestic heat demand peaks.
Also, demand for industrial heat is typically constant so there is potential for fusion (or fission) power plants to supply a post-carbon market with baseload heat.

\subsubsection{Fission-fusion hybrids}

Fission-fusion hybrids\cite{Freidberg2009} are reactors which utilise fissionable material in the blanket of a fusion device, allowing fusion neutrons to trigger non-critical fission reactions to generate additional power \cite{Manheimer2009}.
A key part of this concept is that the fuel need not be ${}^{235}$U, for example it could be recycled ${}^{238}$U waste.

Hybrids would alleviate many significant challenges facing fusion designs.
By reducing the required fusion energy gain, the concomitant heat exhaust loading, reactor size and associated cost are all significantly reduced.
Achieving a tritium-breeding ratio $>1$ becomes significantly more straightforward.
Furthermore, power output would be much higher for the same size reactor vessel, as most of the power production would come from the blanket instead of the plasma \cite{Manheimer2009}.
The higher power output at lower capital cost could make hybrid electricity inherently cheaper than ``pure'' fusion, assuming the increased costs of the non-fusion-specific requirements are low enough. 

Hybridisation would, however, nullify some of the purported advantages of fusion over fission, namely non-proliferation (although this is disputed \cite{Manheimer2014}) and the LLW criterion; hybridisation has thus been generally disregarded.
But if we reconsider its value proposition in light of a relaxed waste criterion, and accept fusion will produce ILW, then it is a much smaller jump to accept that HLW and actinides will also be produced.
But the only significant qualitative difference between a fission-fusion hybrid and a pure-fission device would therefore be the absence of meltdown risk.

In summary, hybrid devices could better compete with next-generation fission through reduced plasma physics complexity, lower LCOE, and inherent safety from meltdown, but this comes at a cost with regards to waste and proliferation concerns.

\subsubsection{Summary}

In a decarbonised grid dominated by cheap intermittent renewables, power generation will likely need to follow demand.
Fusion energy could do this to some extent, which would increase its competitiveness relative to gas with CCS, the technology most economically-suited to fast load-following in a post-carbon world.

There may exist low-carbon heat markets for which fusion is better-suited, but it will not have exclusive access.

\begin{figure}
  \centering
  \includegraphics[width=\textwidth]{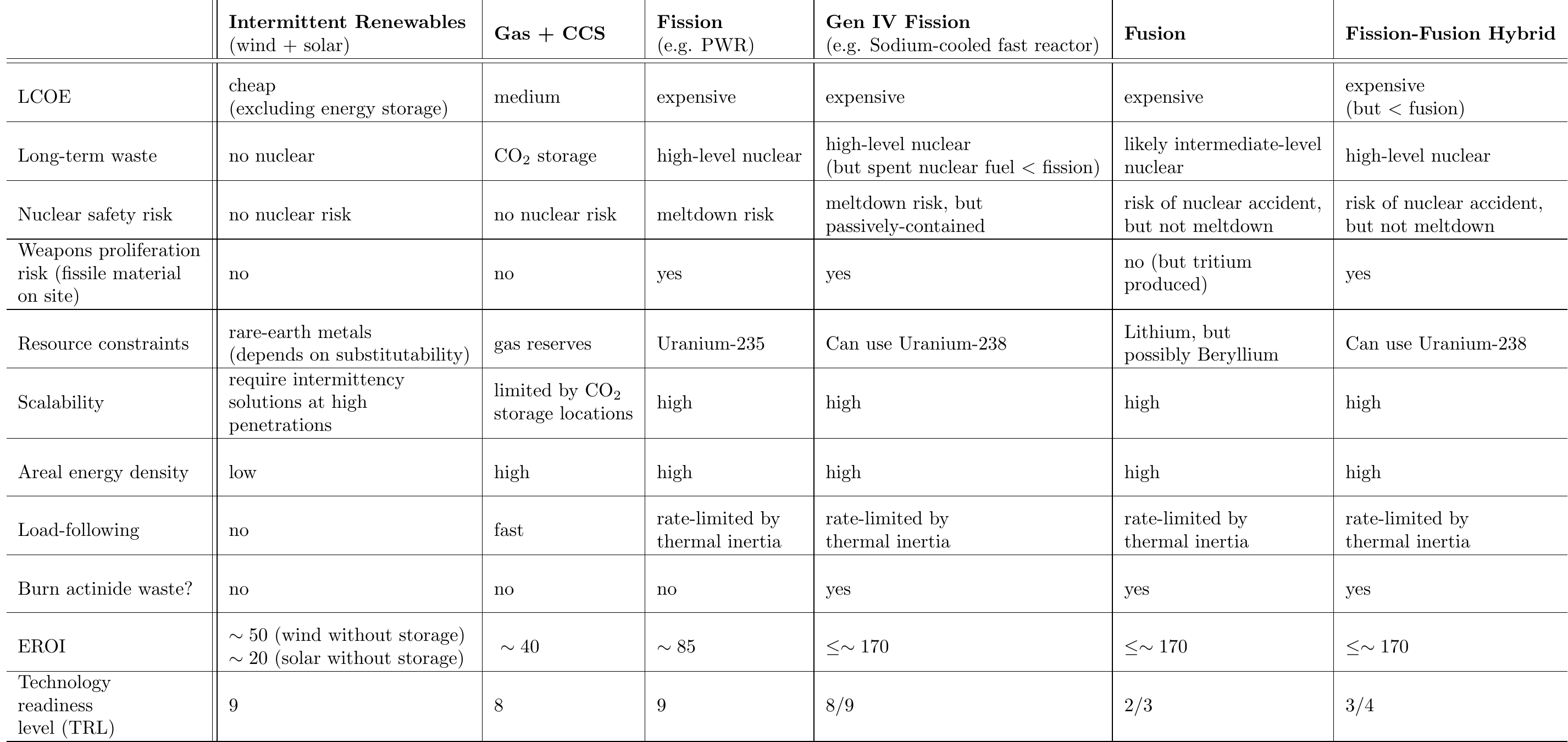}
  \caption{Table of the relative advantages and disadvantages of various energy sources.}
  \label{fig:prosandcons}
\end{figure}

Once realistic engineering and materials constraints are considered, the characteristic advantages of fusion become less clear, and the difference between fusion and fission (especially generation IV fission concepts) becomes finer. We summarize some of this complexity in figure \ref{fig:prosandcons}.

Fusion could potentially achieve a very high EROI, but so could advanced fission fuel cycles, and in both cases the maximum achievable value is significantly constrained by the size of the required power plant.

Given that any practically realisable fusion device is likely to produce nuclear waste which requires deep geological disposal, the step towards a fission-fusion hybrid becomes more acceptable.
This presents the possibility of building an intrinsically meltdown-proof, ${}^{238}$U-burning hybrid reactor, which would not only solve many of the hardest problems with designing a ``pure'' fusion device, but could provide significantly cheaper electricity.

\subsubsection{Sensitivity to assumptions}

Clearly the scenarios presented in this paper are highly speculative, and rely on many assumptions. 
While it’s impossible to be exhaustive, by using our framework we can at least discuss possible implications of changing some of those assumptions:

\begin{itemize}
\item If LLW could be achieved (most likely through regulatory change) then fusion would have a much clearer advantage over fission.

\item If very compact fusion reactors can be developed then they may manage to undercut fission or gas with CCS plants.
However our earlier analysis still indicates that the balance-of-plant required outside the fusion reactor itself might prevent them reaching a lower LCOE than that of future renewables.

\item If very strong unexpected mineral supply constraints apply (such as cartelisation of production of certain rare-earth metals, or refusal to trade uranium internationally) then that could be a limiting factor for deployment of various technologies.

\item If some geopolitical advantage of fusion technology (such as independence or compactness) is prioritized by governments, then that may be a reason for them to pay a premium price.

\item If renewable buildout is stymied such that premise E is violated, it could mean firm sources form the backbone of the decarbonised grid.

\item In a net-zero scenario in which renewables don't reach 100\%, the failure to develop or deploy CCS benefits other low-carbon dispatchable sources such as nuclear considerably.

\item If any of these caveats apply to a specific region, then that region could develop infrastructure following a different pathway.
\end{itemize}

\subsection{What is the current strategy?}
\subsubsection{Public}

The major government-led national and international fusion efforts are pushing towards broadly similar designs: large tokamaks which provide baseload electricity. 

A representative example is given by the Eurofusion roadmap\cite{Donne2018} which emphasises the need for more materials science research ``so as to avoid permanent waste repositories and allow recycling''.
It also generally advocated for baseload because ``a predictable baseload electricity supply is needed to handle short-term and seasonal variations in the renewable sources''.
Furthermore, it states that ``The ultimate goal is commercial electricity'', with no mention of supplying heat directly.

The UK’s recently-announced Spherical Tokamak for Energy Production (STEP) project\cite{STEP2019} will be examining many of these socio-economic and environmental issues in more detail.

\subsubsection{Private}

There are also now many private enterprises ostensibly planning to demonstrate fusion energy production on timelines accelerated relative to the Eurofusion roadmap.

Most of these concepts still have similar characteristics to the conventional government-led programmes’ designs.
Although the confinement technologies being pursued vary widely, the necessity of neutron production and tritium breeding implies these designs will have similar waste production, safety cases, and material supply constraints.

These companies generally plan to achieve net-power output in devices which are much smaller than those which follow the trajectory of the Eurofusion roadmap, with an aim of significantly reducing LCOE by reducing initial capital cost.
Many of these designs rely on recent developments in Rare-Earth Barium Copper Oxide (REBCO) high-temperature superconducting magnet technology in order to scale down the required reactor size\cite{Sykes2018}.
(Essentially following the “raise magnetic field strength'' strategy described by Freidberg\cite{Freidberg2015}.)

The implications of compact reactors are discussed above, but the other significant difference for these enterprises is the projected timelines\cite{Kramer2018, Sykes2018}.
If they can produce a commercially-competitive reactor in time to scale up before decarbonisation, they could displace fossil fuel plants.
This would change the outlook for fusion significantly, but is reliant on very optimistic technological progress.

\section{Conclusions}

There is currently a clear motivation for fusion. 
Renewables alone likely will not present an optimal solution to the energy supply problem\cite{Jenkins2018a}, so grids will benefit from dispatchable low-carbon backup.
Fusion would be one competitor in that market\cite{Bustreo2019}, and has some unique advantages over both current fission and gas with CCS, so even in our simplified scenario continued fusion technology development should be pursued on that basis\footnotemark[1].

\footnotetext[1]{There are also many non-energy-related arguments for the value of fusion research to society, such as spin-offs \cite{Buckingham2016}, multiple-use technology development\cite{Bruzzone2018}, skills training \cite{Reid2014}, interdisciplinary crossover\cite{Diamond2005}, international collaboration \cite{McCray2010}, public science education\cite{MalcolmNeale2019}, and potential high-impulse space propulsion technologies\cite{Santarius2005}.}

However, fusion energy research originally intended to solve a somewhat different set of problems to those which will face a post-carbon grid in the future.
Additionally, the established  relative advantages of fusion energy are somewhat vague, especially surrounding nuclear waste production.
With these extra constraints, fusion is more similar to fission than it first appears; the largest remaining differences are non-criticality and non-proliferation, so these advantages should be pushed while waste is relaxed.

However, in such a renewables-dominated (and market-dominated) scenario, fusion may only end up significantly contributing to post-carbon global energy supply in a scenario where:

\begin{itemize}
\item Renewables and energy storage cannot solve the decarbonisation problem alone,
\item Fusion can help mitigate renewables’ intermittency problems,
\item Or fusion can find a niche market such as baseload industrial or district heating,
\item Production of nuclear waste which requires deep geologic disposal is seen as acceptable,
\item The remaining advantages of fusion over fission are enough to motivate development.
\end{itemize}

If none of these things happen, fusion may be relegated to being a post-CCS or post-Uranium technology, to a much smaller market, or simply never become an established technology.

Therefore fusion programmes should consider:

\begin{enumerate}
\item Seriously studying the feasibility and benefit of the low-level waste criterion, and consider abandoning it, especially for first-of-a-kind devices,
\item Aiming for a dispatchable, load-following reactor design,
\item Supplying heat as well as electricity.
\end{enumerate}

The fusion research community should also explore several additional questions:

\begin{enumerate}
\item What might the full lifecycle Energy Return On Invested be for modern designs of commercial fusion power plants?
\item Which materials and elements currently being considered in fusion prototype designs cannot be scaled to hundreds of GW-sized reactors on sustainability or resource availability grounds?
\item Is the recycling of activated materials from reactor structural materials actually plausible or desirable?
\end{enumerate}

In order to secure the confidence of large-scale private-sector investors, fusion projects will likely need to be able to answer these questions.
However we believe most current conventional fusion reactor concepts and research programmes do not consider these aspects in detail.

\section{Policy Implications}

\begin{itemize}
  \item By the time the government fusion programmes demonstrate fusion energy production, the global energy grid will likely have changed very significantly.
  \item A fusion reactor supplying baseload electricity might be obsolete by the time a demonstration device is built.
  \item The fusion community should consider output power modulation when defining research goals.
  \item Fusion research should consider relaxing the low-level waste criterion to accept intermediate-level waste.
\end{itemize}

\subsection*{Acknowledgements}

The authors gratefully acknowledge extensive discussions with colleagues in multiple fusion research groups, including in ITER and CCFE.

\subsection*{Funding}
The authors of this work are collectively supported through studentships by the Engineering and Physical Sciences Research Council (EP/L01663X/1), a CCFE iCASE studentship, and a Clarendon Scholarship.

\bibliographystyle{elsarticle-num}
\bibliography{Identity} 

\begin{thebibliography}{100}
\expandafter\ifx\csname url\endcsname\relax
  \def\url#1{\texttt{#1}}\fi
\expandafter\ifx\csname urlprefix\endcsname\relax\def\urlprefix{URL }\fi
\expandafter\ifx\csname href\endcsname\relax
  \def\href#1#2{#2} \def\path#1{#1}\fi

\bibitem{Bustreo2019}
C.~Bustreo, U.~Giuliani, D.~Maggio, G.~Zollino,
  \href{https://doi.org/10.1016/j.fusengdes.2019.03.150}{{How fusion power can
  contribute to a fully decarbonized European power mix after 2050}}, Fusion
  Engineering and Design 146~(March) (2019) 2189--2193 (2019).
\newblock \href {https://doi.org/10.1016/j.fusengdes.2019.03.150}
  {\path{doi:10.1016/j.fusengdes.2019.03.150}}.
\newline\urlprefix\url{https://doi.org/10.1016/j.fusengdes.2019.03.150}

\bibitem{Cabal2017}
H.~Cabal, Y.~Lechon, C.~Bustreo, F.~Gracceva, M.~Biberacher, D.~Ward,
  D.~Dongiovanni, P.~E. Gronheit, {Fusion power in a future low carbon global
  electricity system}, Energy Strategy Reviews 15 (2017) 1--8 (2017).
\newblock \href {https://doi.org/10.1016/j.esr.2016.11.002}
  {\path{doi:10.1016/j.esr.2016.11.002}}.

\bibitem{Anyaeji2017}
E.~Anyaeji, {The Economic Impact of Fusion Power in the UK's 2050 Energy Mix},
  Doctor of philosophy, University of Reading (2017).

\bibitem{Keeling1960}
C.~D. Keeling, {The Concentration and Isotopic Abundances of Carbon Dioxide in
  the Atmosphere The Concentration and Isotopic Abundances of Carbon}, Tellus
  2826~(12:2) (1960) 200--203 (1960).
\newblock \href {https://doi.org/10.3402/tellusa.v12i2.9366}
  {\path{doi:10.3402/tellusa.v12i2.9366}}.

\bibitem{Schneider2011}
M.~Schneider, A.~Froggatt, S.~Thomas, {The World Nuclear Industry Status Report
  2010-2011: Nuclear Power in a Post-Fukushima World}, Tech. rep., Worldwatch
  Institute (2011).

\bibitem{IPCC2018}
J.~Rogelj, D.~Shindell, K.~Jiang, S.~Fifita, P.~Forster, V.~Ginzburg, C.~Handa,
  H.~Kheshgi, S.~Kobayashi, E.~Kriegler, L.~Mundaca, R.~S{\'{e}}f{\'{e}}rian,
  M.V.Vilari{\~{n}}o, {Mitigation Pathways Compatible with 1.5C in the Context
  of Sustainable Development}, Tech. rep., Intergovernmental Panel on Climate
  Change (2018).

\bibitem{Kramer2018}
D.~Kramer,
  \href{https://physicstoday.scitation.org/doi/full/10.1063/PT.3.3994}{{Will
  doubling magnetic field strength halve the time to fusion energy ?}}, Physics
  Today 25 (2018) 22--25 (2018).
\newblock \href {https://doi.org/10.1063/PT.3.3994}
  {\path{doi:10.1063/PT.3.3994}}.
\newline\urlprefix\url{https://physicstoday.scitation.org/doi/full/10.1063/PT.3.3994}

\bibitem{Zhuang2019}
G.~Zhuang, G.~Q. Li, J.~Li, Y.~X. Wan, Y.~Liu, X.~L. Wang, Y.~T. Song,
  {Progress of the CFETR design}, Nuclear Fusion 59~(112010) (2019).

\bibitem{Kim2015}
K.~Kim, K.~Im, H.~C. Kim, S.~Oh, J.~S. Park, S.~Kwon, Y.~S. Lee, J.~H. Yeom,
  C.~Lee, G.-S. Lee, G.~Neilson, C.~Kessel, T.~Brown, P.~Titus, D.~Mikkelsen,
  Y.~Zhai, {Design concept of K-DEMO for near-term}, Nuclear Fusion 55 (2015).
\newblock \href {https://doi.org/10.1088/0029-5515/55/5/053027}
  {\path{doi:10.1088/0029-5515/55/5/053027}}.

\bibitem{Federici2017}
G.~Federici, W.~Biel, M.~R. Gilbert, R.~Kemp, N.~Taylor, R.~Wenninger,
  {European DEMO design strategy and consequences for materials}, Nuclear
  Fusion 57~(092002) (2017).

\bibitem{Sorbom2015}
B.~N. Sorbom, J.~Ball, T.~R. Palmer, F.~J. Mangiarotti, J.~M. Sierchio,
  P.~Bonoli, C.~Kasten, D.~A. Sutherland, H.~S. Barnard, C.~B. Haakonsen,
  J.~Goh, C.~Sung, D.~G. Whyte, {ARC: A compact, high-field, fusion nuclear
  science facility and demonstration power plant with demountable magnets},
  Fusion Engineering and Design 100 (2015) 378--405 (2015).
\newblock \href {http://arxiv.org/abs/1409.3540} {\path{arXiv:1409.3540}},
  \href {https://doi.org/10.1016/j.fusengdes.2015.07.008}
  {\path{doi:10.1016/j.fusengdes.2015.07.008}}.

\bibitem{Petty2008}
C.~C. Petty, {Sizing up plasmas using dimensionless parameters}, Physics of
  Plasmas 15~(080501) (2008).
\newblock \href {https://doi.org/10.1063/1.2961043}
  {\path{doi:10.1063/1.2961043}}.

\bibitem{ITERconf1999}
I.~P. E. G. o.~C. and Transport, {Chapter 2 : Plasma confinement and
  transport}, Nuclear Fusion 39~(2175) (1999).

\bibitem{Linsmeier2017}
C.~Linsmeier, M.~Rieth, J.~Aktaa, T.~Chikada, A.~Hoffmann, J.~Hoffmann,
  A.~Houben, H.~Kurishita, X.~Jin, M.~Li, A.~Litnovsky, S.~Matsuo, A.~von
  M{\"{u}}ller, V.~Nikolic, T.~Palacios, R.~Pippan, D.~Qu, J.~Reiser,
  J.~Riesch, T.~Shikama, R.~Stieglitz, T.~Weber, S.~Wurster, J.-H. You,
  Z.~Zhou, {Development of advanced high heat flux and plasma-facing
  materials}, Nuclear Fusion 57~(092007) (2017).

\bibitem{Mukhovatov2007a}
M.~Shimada, D.~Campbell, V.~Mukhovatov, M.~Fujiwara, N.~Kirneva, K.~Lackner,
  M.~Nagami, V.~Pustovitov, N.~Uckan, J.~Wesley, N.~Asakura, A.~Costley,
  A.~Donn{\'{e}}, E.~Doyle, A.~Fasoli, C.~Gormezano, Y.~Gribov, O.~Gruber,
  T.~Hender, W.~Houlberg, S.~Ide, Y.~Kamada, A.~Leonard, B.~Lipschultz,
  A.~Loarte, K.~Miyamoto, V.~Mukhovatov, T.~Osborne, A.~Polevoi, A.~Sips,
  \href{http://iopscience.iop.org/0029-5515/47/6/S01
  http://stacks.iop.org/0029-5515/47/i=6/a=S01?key=crossref.9f5018b333c98316af6e827446ba6d15}{{Chapter
  1: Overview and summary}}, Nuclear Fusion 47~(6) (2007) S1--S17 (2007).
\newblock \href {https://doi.org/10.1088/0029-5515/47/6/S01}
  {\path{doi:10.1088/0029-5515/47/6/S01}}.
\newline\urlprefix\url{http://iopscience.iop.org/0029-5515/47/6/S01
  http://stacks.iop.org/0029-5515/47/i=6/a=S01?key=crossref.9f5018b333c98316af6e827446ba6d15}

\bibitem{Rider1995}
T.~H. Rider, {Fundamental Limitations on Plasma Fusion Systems Not in
  Thermodynamic Equilibrium}, Ph.D. thesis, MIT (1995).

\bibitem{Rider1997}
T.~H. Rider, {Fundamental limitations on plasma fusion systems not in
  thermodynamic equilibrium}, Physics of Plasmas 1039~(1997) (1997) 1039--1046
  (1997).
\newblock \href {https://doi.org/10.1063/1.872556}
  {\path{doi:10.1063/1.872556}}.

\bibitem{Rowcliffe2018}
A.~F. Rowcliffe, L.~M. Garrison, Y.~Yamamoto, L.~Tan, Y.~Katoh,
  \href{http://dx.doi.org/10.1016/j.fusengdes.2017.07.012}{{Materials
  challenges for the fusion nuclear science facility}}, Fusion Engineering and
  Design 135 (2018) 290--301 (2018).
\newblock \href {https://doi.org/10.1016/j.fusengdes.2017.07.012}
  {\path{doi:10.1016/j.fusengdes.2017.07.012}}.
\newline\urlprefix\url{http://dx.doi.org/10.1016/j.fusengdes.2017.07.012}

\bibitem{Federici2017a}
G.~Federici, W.~Biel, M.~R. Gilbert, R.~Kemp, N.~Taylor, R.~Wenninger,
  {European DEMO design strategy and consequences for materials}, Nuclear
  Fusion 57~(9) (2017).
\newblock \href {https://doi.org/10.1088/1741-4326/57/9/092002}
  {\path{doi:10.1088/1741-4326/57/9/092002}}.

\bibitem{Ueda2017}
Y.~Ueda, K.~Schmid, M.~Balden, J.~W. Coenen, T.~Loewenhoff, A.~Ito,
  A.~Hasegawa, C.~Hardie, M.~Porton, M.~Gilbert, {Baseline high heat flux and
  plasma facing materials for fusion}, Nuclear Fusion 57~(9) (2017).
\newblock \href {https://doi.org/10.1088/1741-4326/aa6b60}
  {\path{doi:10.1088/1741-4326/aa6b60}}.

\bibitem{Stork2014}
D.~Stork, P.~Agostini, J.~L. Boutard, D.~Buckthorpe, E.~Diegele, S.~L. Dudarev,
  C.~English, G.~Federici, M.~R. Gilbert, S.~Gonzalez, A.~Ibarra, C.~Linsmeier,
  A.~{Li Puma}, G.~Marbach, P.~F. Morris, L.~W. Packer, B.~Raj, M.~Rieth, M.~Q.
  Tran, D.~J. Ward, S.~J. Zinkle,
  \href{http://dx.doi.org/10.1016/j.jnucmat.2014.06.014}{{Developing
  structural, high-heat flux and plasma facing materials for a near-term DEMO
  fusion power plant: The EU assessment}}, Journal of Nuclear Materials
  455~(1-3) (2014) 277--291 (2014).
\newblock \href {https://doi.org/10.1016/j.jnucmat.2014.06.014}
  {\path{doi:10.1016/j.jnucmat.2014.06.014}}.
\newline\urlprefix\url{http://dx.doi.org/10.1016/j.jnucmat.2014.06.014}

\bibitem{EU2017}
{Working Paper 1976/2016 REV 1: Revised Technical Report on Member States
  Results of the EUCO Policy Scenarios} (2017).

\bibitem{Jenkins2018b}
J.~D. Jenkins, Z.~Zhou, R.~Ponciroli, R.~B. Vilim, F.~Ganda, F.~D. Sisternes,
  A.~Botterud, {The benefits of nuclear flexibility in power system operations
  with renewable energy}, Applied Energy 222 (2018) 872--884 (2018).
\newblock \href {https://doi.org/10.1016/j.apenergy.2018.03.002}
  {\path{doi:10.1016/j.apenergy.2018.03.002}}.

\bibitem{Donne2018}
T.~Donn{\'{e}}, W.~Morris, {European Research Roadmap to the Realisation of
  Fusion Energy}, Tech. rep. (2018).

\bibitem{Cardozo2016}
N.~J.~L. Cardozo, A.~G.~G. Lange, G.~J. Kramer, {Fusion: Expensive and Taking
  Forever?}, Journal of Fusion Energy 35~(1) (2016) 94--101 (2016).
\newblock \href {https://doi.org/10.1007/s10894-015-0012-7}
  {\path{doi:10.1007/s10894-015-0012-7}}.

\bibitem{Surrey2019}
E.~Surrey, {Engineering challenges for accelerated fusion demonstrators},
  Philosophical Transactions of the Royal Society A 377~(2017442) (2019).

\bibitem{Sykes2018}
A.~Sykes, A.~E. Costley, C.~G. Windsor, O.~Asunta, G.~Brittles, P.~Buxton,
  V.~Chuyanov, J.~W. Connor, M.~P. Gryaznevich, B.~Huang, J.~Hugill,
  A.~Kukushkin, D.~Kingham, A.~V. Langtry, S.~Mcnamara, J.~G. Morgan,
  P.~Noonan, J.~S.~H. Ross, V.~Shevchenko, R.~Slade, G.~Smith, {Compact fusion
  energy based on the spherical tokamak}, Nuclear Fusion 58~(016039) (2018).

\bibitem{Carbajales-Dale2012}
M.~Carbajales-dale, C.~J. Barnhart, A.~R. Brandt, S.~M. Benson, {A better
  currency for investing in a sustainable future}, Nature Publishing Group
  (2012).
\newblock \href {https://doi.org/10.1038/nclimate2285}
  {\path{doi:10.1038/nclimate2285}}.

\bibitem{BEIS2016}
{Electricity Generation Costs}, Tech. Rep. November, Department for Business,
  Energy and Industrial Strategy (2016).

\bibitem{McKinsey2019}
C.~Tryggestad, N.~Sharma, O.~Roser, B.~Smeets, J.~van~de Staaij,
  \href{https://www.mckinsey.com/industries/oil-and-gas/our-insights/global-energy-perspective-2019}{{Global
  Energy Perspective 2019 : Reference Case}}, Tech. Rep. January, McKinsey
  (2019).
\newline\urlprefix\url{https://www.mckinsey.com/industries/oil-and-gas/our-insights/global-energy-perspective-2019}

\bibitem{Lazard2019}
{Lazard's Levelized Cost of Energy Analysis—Version 13.0}, Tech. Rep.
  November, Lazard (2019).

\bibitem{IPCCSPM2018}
V.~Masson-Delmotte, P.~Zhai, H.-O. P{\"{o}}rtner, D.~Roberts, J.~Skea,
  P.~Shukla, A.~Pirani, W.~Moufouma-Okia, C.~P{\'{e}}an, R.~Pidcock,
  S.~Connors, J.~Matthews, Y.~Chen, X.~Zhou, M.~Gomis, E.~Lonnoy, T.~Maycock,
  M.~Tignor, T.~Waterfield, {Global Warming of 1.5C. An IPCC Special Report on
  the impacts of global warming of 1.5C above pre-industrial levels and related
  global greenhouse gas emission pathways, in the context of strengthening the
  global response to the threat of climate change,}, Tech. rep., IPCC (2018).

\bibitem{BP2020}
BP,
  \href{https://www.bp.com/content/dam/bp/business-sites/en/global/corporate/pdfs/energy-economics/energy-outlook/bp-energy-outlook-2020.pdf}{{Energy
  Outlook: 2020 edition}}, Tech. rep. (2020).
\newline\urlprefix\url{https://www.bp.com/content/dam/bp/business-sites/en/global/corporate/pdfs/energy-economics/energy-outlook/bp-energy-outlook-2020.pdf}

\bibitem{UKCCC2019}
\href{https://www.theccc.org.uk/publication/net-zero-technical-report/}{{Net
  Zero Technical report}}, Tech. Rep. May, UK Committee on Climate Change
  (2019).
\newline\urlprefix\url{https://www.theccc.org.uk/publication/net-zero-technical-report/}

\bibitem{Entler2018}
S.~Entler, J.~Horacek, T.~Dlouhy, V.~Dostal,
  \href{https://doi.org/10.1016/j.energy.2018.03.130}{{Approximation of the
  economy of fusion energy}}, Energy 152 (2018) 489--497 (2018).
\newblock \href {https://doi.org/10.1016/j.energy.2018.03.130}
  {\path{doi:10.1016/j.energy.2018.03.130}}.
\newline\urlprefix\url{https://doi.org/10.1016/j.energy.2018.03.130}

\bibitem{Sheffield2016}
J.~Sheffield, S.~L. Milora, {Generic Magnetic Fusion Reactor Revisited}, Fusion
  Science and Technology 70~(1) (2016) 14--35 (2016).
\newblock \href {https://doi.org/10.13182/FST15-157}
  {\path{doi:10.13182/FST15-157}}.

\bibitem{Haas2019}
R.~Haas, S.~Thomas, A.~Ajanovic, {The Historical Development of the Costs of
  Nuclear Power}, Springer Fachmedien Wiesbaden, Wiesbaden, 2019, pp. 97--115
  (2019).
\newblock \href {https://doi.org/10.1007/978-3-658-25987-7-5}
  {\path{doi:10.1007/978-3-658-25987-7-5}}.

\bibitem{Jacobson2018}
M.~Z. Jacobson, M.~A. Delucchi, M.~A. Cameron, B.~V. Mathiesen,
  \href{https://doi.org/10.1016/j.renene.2018.02.009}{{Matching demand with
  supply at low cost in 139 countries among 20 world regions with 100 \%
  intermittent wind, water, and sunlight ( WWS ) for all purposes}}, Renewable
  Energy 123 (2018) 236--248 (2018).
\newblock \href {https://doi.org/10.1016/j.renene.2018.02.009}
  {\path{doi:10.1016/j.renene.2018.02.009}}.
\newline\urlprefix\url{https://doi.org/10.1016/j.renene.2018.02.009}

\bibitem{Connolly2014}
D.~Connolly, B.~V. Mathiesen, {A technical and economic analysis of one
  potential pathway to a 100\% renewable energy system}, International Journal
  of Sustainable Energy Planning and Management 1 (2014) 7--28 (2014).
\newblock \href {https://doi.org/10.5278/ijsepm.2014.1.2}
  {\path{doi:10.5278/ijsepm.2014.1.2}}.

\bibitem{Ram2019}
M.~Ram, D.~Bogdanov, A.~Aghahosseini, A.~Gulagi, A.~Oyewo, M.~Child,
  U.~Caldera, K.~Sadovskaia, J.~Farfan, L.~Barbosa, M.~Fasihi, S.~Khalili,
  B.~Dalheimer, G.~Gruber, T.~Traber, F.~{De Caluwe}, H.-J. Fell, C.~Breyer,
  {Global Energy System Based on 100\% Renewable Energy – Power, Heat,
  Transport and Desalination Sectors}, Tech. rep., Lappeenranta University of
  Technology and Energy Watch Group, Berlin (2019).

\bibitem{Ueckerdt2015}
F.~Ueckerdt, R.~Brecha, G.~Luderer,
  \href{http://dx.doi.org/10.1016/j.renene.2015.03.002}{{Analyzing major
  challenges of wind and solar variability in power systems}}, Renewable Energy
  81 (2015) 1--10 (2015).
\newblock \href {https://doi.org/10.1016/j.renene.2015.03.002}
  {\path{doi:10.1016/j.renene.2015.03.002}}.
\newline\urlprefix\url{http://dx.doi.org/10.1016/j.renene.2015.03.002}

\bibitem{Child2019}
M.~Child, C.~Kemfert, D.~Bogdanov, C.~Breyer,
  \href{https://doi.org/10.1016/j.renene.2019.02.077}{{Flexible electricity
  generation , grid exchange and storage for the transition to a 100 \%
  renewable energy system in Europe}}, Renewable Energy 139 (2019) 80--101
  (2019).
\newblock \href {https://doi.org/10.1016/j.renene.2019.02.077}
  {\path{doi:10.1016/j.renene.2019.02.077}}.
\newline\urlprefix\url{https://doi.org/10.1016/j.renene.2019.02.077}

\bibitem{Zappa2019}
W.~Zappa, M.~Junginger, M.~V.~D. Broek, C.~L. Cover,
  \href{https://doi.org/10.1016/j.apenergy.2018.08.109}{{Is a 100 \% renewable
  European power system feasible by 2050 ?}}, Applied Energy 233-234~(July
  2018) (2019) 1027--1050 (2019).
\newblock \href {https://doi.org/10.1016/j.apenergy.2018.08.109}
  {\path{doi:10.1016/j.apenergy.2018.08.109}}.
\newline\urlprefix\url{https://doi.org/10.1016/j.apenergy.2018.08.109}

\bibitem{Pleissmann2017}
G.~Ple{\ss}mann, P.~Blechinger,
  \href{https://doi.org/10.1016/j.energy.2017.03.076}{{Outlook on South-East
  European power system until 2050 : Least-cost decarbonization pathway meeting
  EU mitigation targets}}, Energy 137 (2017) 1041--1053 (2017).
\newblock \href {https://doi.org/10.1016/j.energy.2017.03.076}
  {\path{doi:10.1016/j.energy.2017.03.076}}.
\newline\urlprefix\url{https://doi.org/10.1016/j.energy.2017.03.076}

\bibitem{Gils2017}
H.~C. Gils, Y.~Scholz, T.~Pregger, D.~L.~D. Tena, D.~Heide,
  \href{http://dx.doi.org/10.1016/j.energy.2017.01.115}{{Integrated modelling
  of variable renewable energy-based power supply in Europe}}, Energy 123
  (2017) 173--188 (2017).
\newblock \href {https://doi.org/10.1016/j.energy.2017.01.115}
  {\path{doi:10.1016/j.energy.2017.01.115}}.
\newline\urlprefix\url{http://dx.doi.org/10.1016/j.energy.2017.01.115}

\bibitem{Ziegler2019}
M.~S. Ziegler, M.~Joshua, J.~Song, M.~Ferrara, Y.-m. Chiang, E.~Jessika, M.~S.
  Ziegler, J.~M. Mueller, D.~Pereira, J.~Song, M.~Ferrara, {Storage
  Requirements and Costs of Shaping Renewable Energy Toward Grid
  Decarbonization Storage Requirements and Costs of Shaping Renewable Energy
  Toward Grid Decarbonization}, Joule 3 (2019) 2134--2153 (2019).
\newblock \href {https://doi.org/10.1016/j.joule.2019.06.012}
  {\path{doi:10.1016/j.joule.2019.06.012}}.

\bibitem{Mileva2016}
A.~Mileva, J.~Johnston, J.~H. Nelson, D.~M. Kammen,
  \href{http://dx.doi.org/10.1016/j.apenergy.2015.10.180}{{Power system
  balancing for deep decarbonization of the electricity sector}}, Applied
  Energy 162 (2016) 1001--1009 (2016).
\newblock \href {https://doi.org/10.1016/j.apenergy.2015.10.180}
  {\path{doi:10.1016/j.apenergy.2015.10.180}}.
\newline\urlprefix\url{http://dx.doi.org/10.1016/j.apenergy.2015.10.180}

\bibitem{Sepulveda2018}
N.~A. Sepulveda, D.~Jesse, F.~J. De, R.~K. Lester, N.~A. Sepulveda, J.~D.
  Jenkins, F.~J.~D. Sisternes, R.~K. Lester,
  \href{https://doi.org/10.1016/j.joule.2018.08.006}{{The Role of Firm
  Low-Carbon Electricity Resources in Deep Decarbonization of Power
  Generation}}, Joule 2~(11) (2018) 2403--2420 (2018).
\newblock \href {https://doi.org/10.1016/j.joule.2018.08.006}
  {\path{doi:10.1016/j.joule.2018.08.006}}.
\newline\urlprefix\url{https://doi.org/10.1016/j.joule.2018.08.006}

\bibitem{Brick2016}
S.~Brick, S.~Thernstrom,
  \href{http://dx.doi.org/10.1016/j.tej.2016.03.001}{{Renewables and
  decarbonization : Studies of California , Wisconsin and Germany}}, The
  Electricity Journal 29~(3) (2016) 6--12 (2016).
\newblock \href {https://doi.org/10.1016/j.tej.2016.03.001}
  {\path{doi:10.1016/j.tej.2016.03.001}}.
\newline\urlprefix\url{http://dx.doi.org/10.1016/j.tej.2016.03.001}

\bibitem{Herendeen2019}
R.~Herendeen, \href{https://doi.org/10.1016/j.enpol.2019.04.002}{{Does “100\%
  renewable” trump concern for spatial impacts?}}, Energy Policy 130~(April
  2018) (2019) 304--310 (2019).
\newblock \href {https://doi.org/10.1016/j.enpol.2019.04.002}
  {\path{doi:10.1016/j.enpol.2019.04.002}}.
\newline\urlprefix\url{https://doi.org/10.1016/j.enpol.2019.04.002}

\bibitem{Cany2016}
C.~Cany, C.~Mansilla, G.~Mathonni{\`{e}}re, T.~Duquesnoy, A.~Baschwitz,
  \href{http://dx.doi.org/10.1016/j.enpol.2016.04.037}{{Nuclear and
  intermittent renewables : Two compatible supply options ? The case of the
  French power mix}}, Energy Policy 95 (2016) 135--146 (2016).
\newblock \href {https://doi.org/10.1016/j.enpol.2016.04.037}
  {\path{doi:10.1016/j.enpol.2016.04.037}}.
\newline\urlprefix\url{http://dx.doi.org/10.1016/j.enpol.2016.04.037}

\bibitem{Mechleri2017}
E.~Mechleri, P.~S. Fennell, N.~M. Dowell,
  \href{http://dx.doi.org/10.1016/j.ijggc.2016.09.018}{{Optimisation and
  evaluation of flexible operation strategies for coal- and gas-CCS power
  stations with a multi-period design approach}}, International Journal of
  Greenhouse Gas Control 59 (2017) 24--39 (2017).
\newblock \href {https://doi.org/10.1016/j.ijggc.2016.09.018}
  {\path{doi:10.1016/j.ijggc.2016.09.018}}.
\newline\urlprefix\url{http://dx.doi.org/10.1016/j.ijggc.2016.09.018}

\bibitem{Jenkins2018a}
J.~D. Jenkins, M.~Luke, S.~Thernstrom,
  \href{https://doi.org/10.1016/j.joule.2018.11.013}{{Getting to Zero Carbon
  Emissions in the Electric Power Sector}}, Joule 2~(12) (2018) 2498--2510
  (2018).
\newblock \href {https://doi.org/10.1016/j.joule.2018.11.013}
  {\path{doi:10.1016/j.joule.2018.11.013}}.
\newline\urlprefix\url{https://doi.org/10.1016/j.joule.2018.11.013}

\bibitem{Brouwer2015}
A.~S. Brouwer, M.~V.~D. Broek, A.~Seebregts, A.~Faaij,
  \href{http://dx.doi.org/10.1016/j.apenergy.2015.06.065}{{Operational
  flexibility and economics of power plants in future low-carbon power
  systems}}, Applied Energy 156 (2015) 107--128 (2015).
\newblock \href {https://doi.org/10.1016/j.apenergy.2015.06.065}
  {\path{doi:10.1016/j.apenergy.2015.06.065}}.
\newline\urlprefix\url{http://dx.doi.org/10.1016/j.apenergy.2015.06.065}

\bibitem{Hamacher2013}
T.~Hamacher, M.~Huber, J.~Dorfner, K.~Schaber, A.~M. Bradshaw,
  \href{http://dx.doi.org/10.1016/j.fusengdes.2013.01.074}{{Nuclear fusion and
  renewable energy forms: Are they compatible?}}, Fusion Engineering and Design
  88~(6-8) (2013) 657--660 (2013).
\newblock \href {https://doi.org/10.1016/j.fusengdes.2013.01.074}
  {\path{doi:10.1016/j.fusengdes.2013.01.074}}.
\newline\urlprefix\url{http://dx.doi.org/10.1016/j.fusengdes.2013.01.074}

\bibitem{Maisonnier2005}
D.~Maisonnier, I.~Cook, S.~Pierre, B.~Lorenzo, B.~Edgar, B.~Karin, D.~Pace,
  F.~Robin, G.~Luciano, H.~Stephan, N.~Claudio, N.~Prachai, P.~Aldo, T.~Neill,
  W.~David, {The European power plant conceptual study}, Fusion Engineering and
  Design 75-79 (2005) 1173--1179 (2005).
\newblock \href {https://doi.org/10.1016/j.fusengdes.2005.06.095}
  {\path{doi:10.1016/j.fusengdes.2005.06.095}}.

\bibitem{Ward2015a}
D.~Ward, R.~Kemp, \href{http://dx.doi.org/10.1016/j.fusengdes.2014.11.021}{{The
  resilience of an operating point for a fusion power plant}}, Fusion
  Engineering and Design 98-99 (2015) 2223--2226 (2015).
\newblock \href {https://doi.org/10.1016/j.fusengdes.2014.11.021}
  {\path{doi:10.1016/j.fusengdes.2014.11.021}}.
\newline\urlprefix\url{http://dx.doi.org/10.1016/j.fusengdes.2014.11.021}

\bibitem{Homonnay}
Z.~Homonnay, A.~Hal{\'{a}}csy, Z.~N{\'{e}}meth, S.~Nagy, K.~S{\"{u}}vegh,
  J.~Hayward, {Molten salt energy storage system for DEMO operated in pulsed
  mode}, in: Proceedings of the 2nd IAEA Technical Meeting on First Generation
  of Fusion Power Plants: Design and Technology.

\bibitem{Vanter2012}
N.~Vanter, H.~Ashraf-Ball, J.~Oswald, {Designing Energy Storage for Fusion
  Reactors}, Tech. rep., Oswald Consultancy Ltd. (2012).

\bibitem{Locatelli2015}
G.~Locatelli, S.~Boarin, F.~Pellegrino, M.~E. Ricotti, {Load following with
  Small Modular Reactors (SMR): A real options analysis}, Energy 80 (2015)
  41--54 (2015).
\newblock \href {https://doi.org/10.1016/j.energy.2014.11.040}
  {\path{doi:10.1016/j.energy.2014.11.040}}.

\bibitem{Ponciroli2017}
R.~Ponciroli, Y.~Wang, Z.~Zhou, A.~Botterud, J.~Jenkins, R.~B. Vilim, F.~Ganda,
  Y.~Wang, Z.~Zhou, A.~Botterud, J.~Jenkins, R.~B. Vilim, F.~Ganda,
  R.~Ponciroli,
  \href{https://doi.org/10.1080/00295450.2017.1388668}{{Profitability
  Evaluation of Load-Following Nuclear Units with Physics-Induced Operational
  Constraints}}, Nuclear Technology 200~(3) (2017) 189--207 (2017).
\newblock \href {https://doi.org/10.1080/00295450.2017.1388668}
  {\path{doi:10.1080/00295450.2017.1388668}}.
\newline\urlprefix\url{https://doi.org/10.1080/00295450.2017.1388668}

\bibitem{Domenichini2013}
R.~Domenichini, L.~Mancuso, N.~Ferrari, J.~Davison, {Operating Flexibility of
  Power Plants with Carbon Capture and Storage (CCS)}, Energy Procedia 37
  (2013) 2727--2737 (2013).
\newblock \href {https://doi.org/10.1016/j.egypro.2013.06.157}
  {\path{doi:10.1016/j.egypro.2013.06.157}}.

\bibitem{Alvarez2018}
R.~A. Alvarez, D.~Zavala-Araiza, D.~R. Lyon, D.~T. Allen, Z.~R. Barkley, A.~R.
  Brandt, K.~J. Davis, S.~C. Herndon, D.~J. Jacob, A.~Karion, E.~A. Kort, B.~K.
  Lamb, T.~Lauvaux, J.~D. Maasakkers, A.~J. Marchese, M.~Omara, S.~W. Pacala,
  J.~Peischl, A.~L. Robinson, P.~B. Shepson, C.~Sweeney, A.~Townsend-Small,
  S.~C. Wofsy, S.~P. Hamburg, {Assessment of methane emissions from the U.S.
  oil and gas supply chain}, Science 361~(6398) (2018) 186--188 (2018).
\newblock \href {https://doi.org/10.1126/science.aar7204}
  {\path{doi:10.1126/science.aar7204}}.

\bibitem{Gorley2015}
M.~J. Gorley, {Critical Assessment 12 : Prospects for reduced activation steel
  for fusion plant}, Materials Science and Technology 31~(8) (2015) 975--980
  (2015).
\newblock \href {https://doi.org/10.1179/1743284714Y.0000000732}
  {\path{doi:10.1179/1743284714Y.0000000732}}.

\bibitem{Brager1985}
H.~R. Brager, F.~A. Garner, D.~S. Gelles, M.~L. Hamilton, {Development of
  Reduced Activation Alloys for Fusion Service}, Journal of Nuclear Materials
  134 (1985) 907--911 (1985).

\bibitem{Conn1983}
R.~W. Conn, E.~E. Bloom, J.~W. Davis, R.~E. Gold, R.~Little, K.~R. Schultz,
  D.~L. Smith, F.~W. Wiffen, {Report of the DOE panel on low activation
  materials for fusion applications} (1983).
\newblock \href {https://doi.org/10.2172/7092294} {\path{doi:10.2172/7092294}}.

\bibitem{NDA2017}
\href{https://ukinventory.nda.gov.uk/wp-content/uploads/sites/18/2014/01/2016UKRWMI-UK-Radioactive-waste-inventory-report.pdf}{{Radioactive
  Wastes in the UK : UK Radioactive Waste Inventory Report}}, Tech. rep.,
  Nuclear Decommisioning Authority (2017).
\newline\urlprefix\url{https://ukinventory.nda.gov.uk/wp-content/uploads/sites/18/2014/01/2016UKRWMI-UK-Radioactive-waste-inventory-report.pdf}

\bibitem{Conn1984}
R.~W. Conn, E.~E. Bloom, J.~W. Davis, R.~E. Gold, R.~Little, K.~R. Schultz,
  D.~L. Smith, F.~W. Wiffen,
  \href{https://doi.org/10.13182/FST84-A23106}{{Lower Activation Materials and
  Magnetic Fusion Reactors}}, Nuclear Technology - Fusion 5~(3) (1984) 291--310
  (1984).
\newblock \href {https://doi.org/10.13182/FST84-A23106}
  {\path{doi:10.13182/FST84-A23106}}.
\newline\urlprefix\url{https://doi.org/10.13182/FST84-A23106}

\bibitem{Kohyama1996}
A.~Kohyama, A.~Hishinuma, D.~Gelles, R.~Klueh, W.~Dietz, K.~Ehrlich,
  {Low-activation ferritic and martensitic steels for fusion application},
  Journal of Nuclear Materials 5~(96) (1996).

\bibitem{Klueh2000}
R.~L. Klueh, E.~T. Cheng, M.~L. Grossbeck, E.~E. Bloom, {Impurity effects on
  reduced-activation ferritic steels developed for fusion applications},
  Journal of Nuclear Materials 280 (2000) 353--359 (2000).

\bibitem{Gilbert2018}
M.~R. Gilbert, T.~Eade, C.~Bachmann, U.~Fischer, N.~P. Taylor,
  \href{https://doi.org/10.1016/j.fusengdes.2017.12.019}{{Waste assessment of
  European DEMO fusion reactor designs}}, Fusion Engineering and Design
  136~(August 2017) (2018) 42--48 (2018).
\newblock \href {https://doi.org/10.1016/j.fusengdes.2017.12.019}
  {\path{doi:10.1016/j.fusengdes.2017.12.019}}.
\newline\urlprefix\url{https://doi.org/10.1016/j.fusengdes.2017.12.019}

\bibitem{Lindau2005}
R.~Lindau, A.Moslang, M.~Rieth, M.~Klimiankou, E.~Materna-Morris, A.~Alamob,
  A.-A.~F. Tavassoli, C.~Cayron, A.-M. Lanchad, P.~Fernandez, N.~Baluc,
  R.Schaublin, E.~Diegele, G.~Filacchioni, J.~Rensmanh, B.~Schaafh, E.~Lucon,
  W.~Dietz, {Present development status of EUROFER and ODS-EUROFER for
  application in blanket concepts}, Fusion Engineering and Design 79 (2005)
  989--996 (2005).
\newblock \href {https://doi.org/10.1016/j.fusengdes.2005.06.186}
  {\path{doi:10.1016/j.fusengdes.2005.06.186}}.

\bibitem{Gaganidze2018}
E.~Gaganidze, F.~Gillemot, I.~Szenthe, M.~Gorley, M.~Rieth, E.~Diegele,
  \href{https://doi.org/10.1016/j.fusengdes.2018.06.027}{{Development of
  EUROFER97 database and material property handbook}}, Fusion Engineering and
  Design 135~(July) (2018) 9--14 (2018).
\newblock \href {https://doi.org/10.1016/j.fusengdes.2018.06.027}
  {\path{doi:10.1016/j.fusengdes.2018.06.027}}.
\newline\urlprefix\url{https://doi.org/10.1016/j.fusengdes.2018.06.027}

\bibitem{Gilbert2019}
M.~R. Gilbert, T.~Eade, T.~Rey, R.~Vale, C.~Bachmann, U.~Fischer, {Waste
  implications from minor impurities in European DEMO materials}, Nuclear
  Fusion 59~(076015) (2019).

\bibitem{Bailey2020}
G.~W. Bailey, M.~R. Gilbert, O.~Vilkhivskaya, {Waste Classification Assessment
  of Nuclear Steels for Fusion Power Applications}, PHYSOR 2020: Transition to
  a Scalable Nuclear Future (2020).

\bibitem{Reid2020}
J.~Reid, G.~Bailey, E.~Cracknell, M.~Gilbert, L.~Packer, {Comparision of waste
  due to irradiated steels in the ESFR and DEMO}, PHYSOR 2020: Transition to a
  Scalable Nuclear Future (2020).

\bibitem{Someya2015}
Y.~Someya, K.~Tobita, H.~Utoh, N.~Asakura, K.~Hoshino, M.~Nakamura,
  S.~Tokunaga, {Management Strategy for Radioactive Waste in the Fusion DEMO
  Reactor}, Fusion Science and Technology 68~(2) (2015) 423--427 (2015).
\newblock \href {https://doi.org/10.13182/FST15-101}
  {\path{doi:10.13182/FST15-101}}.

\bibitem{Pace2012}
L.~D. Pace, L.~El-guebaly, B.~Kolbasov, V.~Massaut, M.~Zucchetti,
  \href{http://www.intechopen.com/books/radioactive-waste/radioactive-
  waste-management-of-fusion-power-plants}{{Radioactive Waste Management of
  Fusion Power Plants}}, in: D.~R.~A. Rahman (Ed.), Radioactive Waste, InTech,
  2012, Ch. Radioactiv (2012).
\newline\urlprefix\url{http://www.intechopen.com/books/radioactive-waste/radioactive-
  waste-management-of-fusion-power-plants}

\bibitem{Goldston2012}
R.~J. Goldston, A.~Glaser, {Safeguard Requirements for Fusion Power Plants},
  Tech. rep. (2012).

\bibitem{GenIVRoadmap2014}
C.~Behar,
  \href{https://www.gen-4.org/gif/upload/docs/application/pdf/2014-03/gif-tru2014.pdf}{{Technology
  Roadmap Update for Generation IV Nuclear Energy Systems}}, Tech. rep., OECD
  Nuclear Energy Agency (2014).
\newline\urlprefix\url{https://www.gen-4.org/gif/upload/docs/application/pdf/2014-03/gif-tru2014.pdf}

\bibitem{IAEA1980}
IAEA, {The Convention on the physical protection of nuclear material} (1980).

\bibitem{Hesketh2010}
{Kevin Hesketh}, A.~Worral, {The Thorium Fuel Cycle}, Tech. Rep. August, UK
  National Nuclear Laboratory (2010).

\bibitem{Raugei2016}
M.~Raugei, E.~Leccisi, \href{http://dx.doi.org/10.1016/j.enpol.2015.12.011}{{A
  comprehensive assessment of the energy performance of the full range of
  electricity generation technologies deployed in the United Kingdom}}, Energy
  Policy 90 (2016) 46--59 (2016).
\newblock \href {https://doi.org/10.1016/j.enpol.2015.12.011}
  {\path{doi:10.1016/j.enpol.2015.12.011}}.
\newline\urlprefix\url{http://dx.doi.org/10.1016/j.enpol.2015.12.011}

\bibitem{White2000}
S.~W. White, G.~L. Kulcinski, {Birth to death analysis of the energy payback
  ratio and CO2 gas emission rates from coal, fission, wind, and DT-fusion
  electrical power plants}, Fusion Engineering and Design 48~(3) (2000)
  473--481 (2000).
\newblock \href {https://doi.org/10.1016/S0920-3796(00)00158-7}
  {\path{doi:10.1016/S0920-3796(00)00158-7}}.

\bibitem{Lozada-Hidalgo2017}
M.~Lozada-Hidalgo, S.~Zhang, S.~Hu, A.~Esfandiar, I.~V. Grigorieva, A.~K. Geim,
  {Scalable and efficient separation of hydrogen isotopes using graphene-based
  electrochemical pumping}, Nature Communications 8~(May) (2017) 1--5 (2017).
\newblock \href {https://doi.org/10.1038/ncomms15215}
  {\path{doi:10.1038/ncomms15215}}.

\bibitem{Rae1978}
H.~K. RAE, {Selecting Heavy Water Processes}, in: Separation of Hydrogen
  Isotopes, 1978, Ch.~1, pp. 1--26 (1978).
\newblock \href {https://doi.org/10.1021/bk-1978-0068.ch001}
  {\path{doi:10.1021/bk-1978-0068.ch001}}.

\bibitem{Weissbach2013}
D.~Wei{\ss}bach, G.~Ruprecht, A.~Huke, K.~Czerski, S.~Gottlieb, A.~Hussein,
  \href{http://dx.doi.org/10.1016/j.energy.2013.01.029}{{Energy intensities ,
  EROIs ( energy returned on invested ), and energy payback times of
  electricity generating power plants}}, Energy 52 (2013) 210--221 (2013).
\newblock \href {https://doi.org/10.1016/j.energy.2013.01.029}
  {\path{doi:10.1016/j.energy.2013.01.029}}.
\newline\urlprefix\url{http://dx.doi.org/10.1016/j.energy.2013.01.029}

\bibitem{Freidberg2015}
J.~P. Freidberg, F.~J. Mangiarotti, J.~Minervini,
  \href{http://dx.doi.org/10.1063/1.4923266}{{Designing a tokamak fusion
  reactor — How does plasma physics fit in ?}}, Physics of Plasmas
  070901~(22) (2015).
\newblock \href {https://doi.org/10.1063/1.4923266}
  {\path{doi:10.1063/1.4923266}}.
\newline\urlprefix\url{http://dx.doi.org/10.1063/1.4923266}

\bibitem{Carbajales-dale2014}
M.~Carbajales-Dale, C.~J. Barnhart, S.~M. Benson, {Can we afford storage? A
  dynamic net energy analysis of renewable electricity generation supported by
  energy storage}, Energy and Environmental Science 7~(5) (2014) 1538--1544
  (2014).
\newblock \href {https://doi.org/10.1039/c3ee42125b}
  {\path{doi:10.1039/c3ee42125b}}.

\bibitem{Fasel2005}
D.~Fasel, M.~Q. Tran, {Availability of lithium in the context of future D – T
  fusion reactors}, Fusion Engineering and Design 79 (2005) 1163--1168 (2005).
\newblock \href {https://doi.org/10.1016/j.fusengdes.2005.06.345}
  {\path{doi:10.1016/j.fusengdes.2005.06.345}}.

\bibitem{JaksulaLi2020}
B.~Jaksula,
  \href{https://pubs.usgs.gov/periodicals/mcs2020/mcs2020-lithium.pdf}{{Mineral
  Commodity Summaries: Lithium}}, Tech. rep., U.S. Geological Survey (2020).
\newline\urlprefix\url{https://pubs.usgs.gov/periodicals/mcs2020/mcs2020-lithium.pdf}

\bibitem{Bradshaw2011}
A.~M. Bradshaw, T.~Hamacher, U.~Fischer,
  \href{http://dx.doi.org/10.1016/j.fusengdes.2010.11.040}{{Is nuclear fusion a
  sustainable energy form?}}, Fusion Engineering and Design 86~(9-11) (2011)
  2770--2773 (2011).
\newblock \href {https://doi.org/10.1016/j.fusengdes.2010.11.040}
  {\path{doi:10.1016/j.fusengdes.2010.11.040}}.
\newline\urlprefix\url{http://dx.doi.org/10.1016/j.fusengdes.2010.11.040}

\bibitem{IEA2019}
\href{https://www.iea.org/reports/global-ev-outlook-2019}{{Global EV Outlook
  2019}}, Tech. rep., IEA (2019).
\newline\urlprefix\url{https://www.iea.org/reports/global-ev-outlook-2019}

\bibitem{Bardi2010}
U.~Bardi, D.~Chimica, U.~Firenze, {Extracting Minerals from Seawater: An Energy
  Analysis}, Sustainability 2 (2010) 980--992 (2010).
\newblock \href {https://doi.org/10.3390/su2040980}
  {\path{doi:10.3390/su2040980}}.

\bibitem{RedBook2003}
\href{http://www.oecd-nea.org/ndd/pubs/2004/5291-uranium-2003.pdf}{{Uranium
  2003: Resources, Production and Demand}}, Tech. rep., OECD Nuclear Energy
  Agency (2004).
\newline\urlprefix\url{http://www.oecd-nea.org/ndd/pubs/2004/5291-uranium-2003.pdf}

\bibitem{Schneider2008}
E.~A. Schneider, W.~C. Sailor, {Long-term uranium supply estimates}, Nuclear
  Technology 162~(3) (2008) 379--387 (2008).
\newblock \href {https://doi.org/10.13182/NT08-A3963}
  {\path{doi:10.13182/NT08-A3963}}.

\bibitem{Schaffer2013}
M.~B. Schaffer, \href{http://dx.doi.org/10.1016/j.enpol.2013.04.062}{{Abundant
  thorium as an alternative nuclear fuel Important waste disposal and weapon
  proliferation advantages}}, Energy Policy 60 (2013) 4--12 (2013).
\newblock \href {https://doi.org/10.1016/j.enpol.2013.04.062}
  {\path{doi:10.1016/j.enpol.2013.04.062}}.
\newline\urlprefix\url{http://dx.doi.org/10.1016/j.enpol.2013.04.062}

\bibitem{Alexandratos2016}
S.~D. Alexandratos, S.~Kung, {Preface to the special issue: Uranium in
  seawater}, Industrial and Engineering Chemistry Research 55~(15) (2016)
  4101--4102 (2016).
\newblock \href {https://doi.org/10.1021/acs.iecr.6b01293}
  {\path{doi:10.1021/acs.iecr.6b01293}}.

\bibitem{Gabriel2013}
S.~Gabriel, A.~Baschwitz, G.~Mathonni{\`{e}}re, T.~Eleouet, F.~Fizaine,
  \href{http://dx.doi.org/10.1016/j.anucene.2013.03.010}{{A critical assessment
  of global uranium resources, including uranium in phosphate rocks, and the
  possible impact of uranium shortages on nuclear power fleets}}, Annals of
  Nuclear Energy 58 (2013) 213--220 (2013).
\newblock \href {https://doi.org/10.1016/j.anucene.2013.03.010}
  {\path{doi:10.1016/j.anucene.2013.03.010}}.
\newline\urlprefix\url{http://dx.doi.org/10.1016/j.anucene.2013.03.010}

\bibitem{Budinis2018}
S.~Budinis, S.~Krevor, N.~M. Dowell, N.~Brandon, A.~Hawkes,
  \href{https://doi.org/10.1016/j.esr.2018.08.003}{{An assessment of CCS costs
  , barriers and potential}}, Energy Strategy Reviews 22~(May) (2018) 61--81
  (2018).
\newblock \href {https://doi.org/10.1016/j.esr.2018.08.003}
  {\path{doi:10.1016/j.esr.2018.08.003}}.
\newline\urlprefix\url{https://doi.org/10.1016/j.esr.2018.08.003}

\bibitem{JaksulaBe2020}
B.~Jaksula,
  \href{https://pubs.usgs.gov/periodicals/mcs2020/mcs2020-beryllium.pdf}{{Mineral
  Commodity Summaries: Beryllium}}, Tech. rep., U.S. Geological Survey (2020).
\newline\urlprefix\url{https://pubs.usgs.gov/periodicals/mcs2020/mcs2020-beryllium.pdf}

\bibitem{Malang2011}
S.~Malang, M.~Tillack, C.~P.~C. Wong, N.~Morley, S.~Smolentsev, {Development of
  the Lead Lithium (DCLL) Blanket Concept}, Fusion Science and Technology
  60~(1) (2011) 249--256 (2011).
\newblock \href {https://doi.org/10.13182/fst10-212}
  {\path{doi:10.13182/fst10-212}}.

\bibitem{Abdou2015}
M.~Abdou, N.~B. Morley, S.~Smolentsev, A.~Ying, S.~Malang, A.~Rowcliffe,
  M.~Ulrickson,
  \href{http://dx.doi.org/10.1016/j.fusengdes.2015.07.021}{{Blanket/first wall
  challenges and required R\&D on the pathway to DEMO}}, Fusion Engineering and
  Design 100 (2015) 2--43 (2015).
\newblock \href {https://doi.org/10.1016/j.fusengdes.2015.07.021}
  {\path{doi:10.1016/j.fusengdes.2015.07.021}}.
\newline\urlprefix\url{http://dx.doi.org/10.1016/j.fusengdes.2015.07.021}

\bibitem{Jun2020}
J.~Jun, P.~F. Tortorelli,
  \href{http://www.sciencedirect.com/science/article/pii/B9780128035818116273}{{Corrosion
  in Other Liquid Metals (Li, PbLi, Hg, Sn, Ga)}}, in: Reference Module in
  Materials Science and Materials Engineering, Elsevier, 2020 (2020).
\newblock \href
  {https://doi.org/https://doi.org/10.1016/B978-0-12-803581-8.11627-3}
  {\path{doi:https://doi.org/10.1016/B978-0-12-803581-8.11627-3}}.
\newline\urlprefix\url{http://www.sciencedirect.com/science/article/pii/B9780128035818116273}

\bibitem{Bruzzone2018}
P.~Bruzzone, W.~H. Fietz, J.~V. Minervini, M.~Novikov, N.~Yanagi, Y.~Zhai,
  J.~Zheng, {High temperature superconductors for fusion magnets}, Nuclear
  Fusion 58~(10) (2018).
\newblock \href {https://doi.org/10.1088/1741-4326/aad835}
  {\path{doi:10.1088/1741-4326/aad835}}.

\bibitem{Bradshaw2013}
A.~M. Bradshaw, T.~Hamacher,
  \href{http://dx.doi.org/10.1016/j.fusengdes.2013.01.059}{{Nuclear fusion and
  the helium supply problem}}, Fusion Engineering and Design 88~(9-10) (2013)
  2694--2697 (2013).
\newblock \href {https://doi.org/10.1016/j.fusengdes.2013.01.059}
  {\path{doi:10.1016/j.fusengdes.2013.01.059}}.
\newline\urlprefix\url{http://dx.doi.org/10.1016/j.fusengdes.2013.01.059}

\bibitem{Tao2011}
C.~S. Tao, J.~Jiang, M.~Tao,
  \href{http://dx.doi.org/10.1016/j.solmat.2011.06.013}{{Natural resource
  limitations to terawatt-scale solar cells}}, Solar Energy Materials and Solar
  Cells 95~(12) (2011) 3176--3180 (2011).
\newblock \href {https://doi.org/10.1016/j.solmat.2011.06.013}
  {\path{doi:10.1016/j.solmat.2011.06.013}}.
\newline\urlprefix\url{http://dx.doi.org/10.1016/j.solmat.2011.06.013}

\bibitem{Kleijn2011}
R.~Kleijn, E.~V.~D. Voet, G.~Jan, L.~V. Oers, C.~V.~D. Giesen,
  \href{http://dx.doi.org/10.1016/j.energy.2011.07.003}{{Metal requirements of
  low-carbon power generation}}, Energy 36~(9) (2011) 5640--5648 (2011).
\newblock \href {https://doi.org/10.1016/j.energy.2011.07.003}
  {\path{doi:10.1016/j.energy.2011.07.003}}.
\newline\urlprefix\url{http://dx.doi.org/10.1016/j.energy.2011.07.003}

\bibitem{IEO2019}
\href{www.eia.gov/ieo}{{International Energy Outlook 2019}}, Tech. rep., U. S.
  Energy Information Administration (2019).
\newline\urlprefix\url{www.eia.gov/ieo}

\bibitem{Thompson2013}
V.~Thompson,
  \href{https://www.gov.uk/government/statistics/energy-trends-september-2013-special-feature-articles-estimates-of-heat-use-in-the-united-kingdom-in-2012}{{Energy
  Trends: September 2013, special feature articles - Estimates of heat use in
  the United Kingdom in 2012}}, Tech. Rep. September, Department of Energy and
  Climate Change (2013).
\newline\urlprefix\url{https://www.gov.uk/government/statistics/energy-trends-september-2013-special-feature-articles-estimates-of-heat-use-in-the-united-kingdom-in-2012}

\bibitem{McMillan2018}
C.~McMillan, V.~Narwade, \href{https://dx.doi.org/10.7799/1481899}{{NREL Data:
  United States County-Level Industrial Energy Use}}, Tech. rep., National
  Renewable Energy Laboratory (2018).
\newblock \href {https://doi.org/10.7799/1481899} {\path{doi:10.7799/1481899}}.
\newline\urlprefix\url{https://dx.doi.org/10.7799/1481899}

\bibitem{McMillan2019a}
C.~McMillan, M.~Ruth, \href{https://dx.doi.org/10.7799/1461488}{{NREL Data:
  Industrial Process Heat Demand Characterization}}, Tech. rep., National
  Renewable Energy Laboratory (2019).
\newblock \href {https://doi.org/10.7799/1461488} {\path{doi:10.7799/1461488}}.
\newline\urlprefix\url{https://dx.doi.org/10.7799/1461488}

\bibitem{Tobita2018}
K.~Tobita, R.~Hiwatari, H.~Utoh, Y.~Miyoshi, N.~Asakura,
  \href{https://doi.org/10.1016/j.fusengdes.2018.04.059}{{Overview of the DEMO
  conceptual design activity in Japan}}, Fusion Engineering and Design
  136~(April) (2018) 1024--1031 (2018).
\newblock \href {https://doi.org/10.1016/j.fusengdes.2018.04.059}
  {\path{doi:10.1016/j.fusengdes.2018.04.059}}.
\newline\urlprefix\url{https://doi.org/10.1016/j.fusengdes.2018.04.059}

\bibitem{Wilson2013}
I.~A.~G. Wilson, A.~J.~R. Rennie, Y.~Ding, P.~C. Eames, P.~J. Hall, N.~J.
  Kelly, \href{http://dx.doi.org/10.1016/j.enpol.2013.05.110}{{Historical daily
  gas and electrical energy flows through Great Britain's transmission networks
  and the decarbonisation of domestic heat}}, Energy Policy 61 (2013) 301--305
  (2013).
\newblock \href {https://doi.org/10.1016/j.enpol.2013.05.110}
  {\path{doi:10.1016/j.enpol.2013.05.110}}.
\newline\urlprefix\url{http://dx.doi.org/10.1016/j.enpol.2013.05.110}

\bibitem{Freidberg2009}
J.~Freidberg, {Research Needs for Fusion-Fission Hybrid Systems: Report of the
  Research Needs Workshop (ReNeW) Gaithersburg, Maryland Sept 30 – Oct 2,
  2009}, Tech. rep., US Department of Energy (2009).

\bibitem{Manheimer2009}
W.~Manheimer, {Hybrid Fusion: The Only Viable Development Path for Tokamaks?},
  Journal of Fusion Energy 28 (2009) 60--82 (2009).
\newblock \href {https://doi.org/10.1007/s10894-008-9156-z}
  {\path{doi:10.1007/s10894-008-9156-z}}.

\bibitem{Manheimer2014}
W.~Manheimer, {Fusion Breeding for Mid-Century Sustainable Power}, Journal of
  Fusion Energy 33 (2014) 199--234 (2014).
\newblock \href {https://doi.org/10.1007/s10894-014-9690-9}
  {\path{doi:10.1007/s10894-014-9690-9}}.

\bibitem{STEP2019}
{U.K. Atomic Energy Authority},
  \href{https://ccfe.ukaea.uk/uk-to-take-a-big-step-to-fusion-electricity/}{{U.K.
  to Take a Big ‘STEP' to Fusion Electricity}} (2019).
\newline\urlprefix\url{https://ccfe.ukaea.uk/uk-to-take-a-big-step-to-fusion-electricity/}

\bibitem{Buckingham2016}
R.~Buckingham, A.~Loving, {Remote-handling challenges in fusion research and
  beyond}, Nature Physics 12~(5) (2016) 391--393 (2016).
\newblock \href {https://doi.org/10.1038/nphys3755}
  {\path{doi:10.1038/nphys3755}}.

\bibitem{Reid2014}
G.~Reid, \href{https://www.ncub.co.uk/reports/why-science.html}{{Why should the
  taxpayer fund science and research?}}, Tech. Rep. December, National Centre
  for Universities and Businesses (2014).
\newline\urlprefix\url{https://www.ncub.co.uk/reports/why-science.html}

\bibitem{Diamond2005}
P.~H. Diamond, S.~I. Itoh, K.~Itoh, T.~S. Hahm, {Zonal flows in plasma - A
  review}, Plasma Physics and Controlled Fusion 47~(5) (2005).
\newblock \href {https://doi.org/10.1088/0741-3335/47/5/R01}
  {\path{doi:10.1088/0741-3335/47/5/R01}}.

\bibitem{McCray2010}
W.~P. Mccray, {'Globalization with hardware': ITER's fusion of technology,
  policy, and politics}, History and Technology 26~(4) (2010) 283--312 (2010).
\newblock \href {https://doi.org/10.1080/07341512.2010.523171}
  {\path{doi:10.1080/07341512.2010.523171}}.

\bibitem{MalcolmNeale2019}
A.~Malcolm-Neale, {Using the wider science curriculum to investigate fusion
  energy}, Physics Education 54~(4) (2019) 044001 (2019).
\newblock \href {https://doi.org/10.1088/1361-6552/ab129a}
  {\path{doi:10.1088/1361-6552/ab129a}}.

\bibitem{Santarius2005}
J.~Santarius, \href{http://fti.neep.wisc.edu/pdf/fdm1287.pdf}{{Fusion Space
  Propulsion – A Shorter Time Frame Than You Think}}, in: Presented at the
  53rd Joint Army-Navy-NASA-Air Force (JANNAF) Propulsion Meeting, Monterey,
  5-8 December 2005., 2005 (2005).
\newline\urlprefix\url{http://fti.neep.wisc.edu/pdf/fdm1287.pdf}

\end{thebibliography}

\end{document}